\documentclass[english,pra,preprint,nofootinbib,superscriptaddress]{revtex4-1}
\usepackage[T1]{fontenc}
\usepackage[latin9]{inputenc}
\usepackage{geometry}
\usepackage{color}
\usepackage{babel}
\usepackage{amsmath}
\usepackage{amssymb}
\usepackage{graphicx}
\usepackage{esint}
\usepackage[unicode=true,pdfusetitle,
 bookmarks=true,bookmarksnumbered=false,bookmarksopen=false,
 breaklinks=false,pdfborder={0 0 0},backref=false,colorlinks=true]
 {hyperref}
 
\begin{document}
\global\long\def\x{\times}
\global\long\def\t{\cdot}
\global\long\def\d{\mathrm{d}}
\global\long\def\ket#1{\left|#1\right\rangle }
\global\long\def\bra#1{\left\langle #1\right|}
\global\long\def\braket#1#2{\langle#1|#2\rangle}
\global\long\def\braoket#1#2#3{\left\langle #1\middle\vert#2\middle\vert#3\right\rangle }
\global\long\def\i{\mathrm{i}}
\global\long\def\e{\mathrm{e}}

\title{Propensity rules in photoelectron circular dichroism in chiral molecules 
I: Chiral hydrogen}

\author{Andres F. Ordonez}
\email{ordonez@mbi-berlin.de}
\affiliation{Max-Born-Institut, Berlin, Germany}
\affiliation{Technische Universit\"at Berlin, Berlin, Germany}
\author{Olga Smirnova}
\email{smirnova@mbi-berlin.de}
\affiliation{Max-Born-Institut, Berlin, Germany}
\affiliation{Technische Universit\"at Berlin, Berlin, Germany}

\begin{abstract}
Photoelectron circular dichroism results from one-photon ionization
of chiral molecules by circularly polarized light and manifests itself
in forward-backward asymmetry of electron emission in the direction
orthogonal to the light polarization plane. What is the physical mechanism
underlying asymmetric electron ejection? How ``which
way'' information builds up in a chiral molecule and
maps into forward-backward asymmetry? 
%In this and in a companion paper \cite{ordonez_2018_alignment} %we identify and intersect the geometrical and dynamical origins of chiral response in photoionization.

%To address its geometrical origin we establish a rigorous relation between the
%responses of unaligned and partially or perfectly aligned molecules \cite{ordonez_2018_alignment}.
%To address its dynamical origin
We introduce
instances of bound chiral wave functions resulting from stationary
superpositions of states in a hydrogen atom and use them to show that the chiral
response in one-photon ionization of aligned molecular ensembles originates from two propensity rules:
(i) Sensitivity of ionization to the sense of electron rotation in the
polarization plane. (ii) Sensitivity of ionization to the direction
of charge displacement or stationary current orthogonal to the polarization
plane. %The interference of two ionization pathways involving these
%selection rules is necessary to observe photoelectron circular dichroism.
In the companion paper \cite{ordonez_2018_alignment} we show 
how the ideas presented here are part of a broader picture valid for all chiral molecules and arbitrary degrees of molecular alignment.
%why our conclusions derived for exotic chiral states
%remain valid for arbitrary chiral molecules.%, and derive general expressions
%for photoelectron circular dichroism in the presence of molecular alignment. 
\end{abstract}
\maketitle

\section{Introduction}

Photoelectron circular dichroism (PECD) \cite{ritchie_theory_1976, powis00, bowering_asymmetry_2001}
heralded the ``dipole revolution''
in chiral discrimination: chiral discrimination without using chiral
light. PECD belongs to a family of methods exciting rotational \cite{patterson_enantiomer-specific_2013,patterson_sensitive_2013,yurchenko_2016,patterson_2017},
electronic, and vibronic \cite{fischer2001isotropic,beaulieu_PXCD}
chiral dynamics without relying on relatively weak interactions with magnetic
fields. In all these methods the chiral response arises already in the electric-dipole approximation
and is significantly higher than in conventional techniques, such
as e.g. absorption circular dichroism or optical rotation, known since
the XIX century (see e.g. \cite{condon_theories_1937}). The connection
between these electric-dipole-approximation-based methods is analyzed in \cite{ordonez_generalized_2018}.
The key feature that distinguishes them from standard techniques is that chiral discrimination relies on a chiral observer - the chiral reference frame defined  by the electric field vectors and detector axis \cite{ordonez_generalized_2018}. 
In PECD, ionization with circularly polarized light of a non-racemic
mixture of randomly-oriented chiral molecules results in a forward-backward
asymmetry (FBA) in the photoelectron angular distribution and
is a very sensitive probe of photoionization dynamics and of molecular
structure and conformation \cite{powis_photoelectron_2008,nahon_valence_2015}.
PECD yields a chiral response as high as few tens of percent of
the total signal and the method is quickly expanding from the realm
of fundamental research to innovative applications, becoming a new
tool in analytical chemistry \cite{fanood2015enantiomer,boesl2016mass,nahon2018new}.
PECD is studied extensively both experimentally \cite{bowering_asymmetry_2001, garcia_2003, turchini_2004, Hergenhahn_2004, Lischke_2004, Stranges_2005, Giardini_2005, Harding_2005, nahon_determination_2006, Contini_2007, Garcia_2008, ulrich_giant_2008, Powis_2008_CPC_9, Powis_2008_PRA_78, Turchini_2009, Garcia_2010, Nahon_2010, Daly_2011, Catone_2012, Catone_2013, garcia13, Turchini_2013, Tia_2013, Powis_2014, Tia_2014, Nahon2016_det, Garcia_2016, Catone_2017} and theoretically
\cite{ritchie_theory_1976, cherepkov_circular_1982, powis00, Powis_lAlanine_2000, stener_density_2004, Harding_2006, Tommaso_2006, stener2006theoretical,  dreissigacker_photoelectron_2014, Artemyev_2015, Koch2017, Ilchen_2017, Tia_2017, Daly_2017, Miles_2017, ordonez_generalized_2018}
and was recently pioneered in the multiphoton \cite{lux_circular_2012, lehmann_imaging_2013, Rafiee_2014, Rafiee_2015, lux_photoelectron_2015,  Lux_2016, rafiee2016wavelength, Kastner_2016, beaulieu_science_2017, Kastner_2017},
pump-probe \cite{comby_relaxation_2016}, and strong-field ionization
regimes \cite{beaulieu_universality_2016, Beaulieu_2016_Faraday}.

%SOME WORDS ABOUT ALIGNMENT 

In this work we focus on the physical mechanisms underlying the chiral response
in one-photon ionization at the level of electrons and introduce ``elementary chiral instances''
- chiral electronic wave functions of the hydrogen atom. 

In molecules, with the exception of the ground electronic state, the
chiral configuration of the nuclei is not a prerequisite for obtaining
a chiral electronic wave function. Thus, one may consider using a
laser field to imprint chirality on the electronic wave function of
an achiral nuclear configuration. The ability to create a chiral electronic
wave function in an atom via a chiral laser field \cite{ayuso_locally_2018} implies the possibility
of creating perfectly oriented (and even stationary) ensembles of
\emph{synthetic chiral molecules} (atoms with chiral electronic wave
functions) with a well defined handedness in a time-resolved fashion
from an initially isotropic ensemble of atoms. Such time-resolved chiral 
control may open new possibilities in the fields of enantiomeric recognition
and enrichment if the ensemble of synthetic chiral atoms 
is made to interact with actual chiral molecules. From a more fundamental point of view, the elementary chiral instances could
be excited in atoms arranged in a lattice of arbitrary symmetry to
explore an interplay of electronic chirality and lattice symmetry
possibly leading to interesting synthetic chiral phases of matter. 

% We first introduce three different classes of chiral bound states.
% The first class is characterized by the presence of a permanent dipole
% and a planar stationary current. The second class is characterized
% by the presence of a stationary helical current. The third class of
% states does not possess any currents, but is characterized by a helical
% electron density which can be decomposed into two helical currents flowing in opposite directions. In all cases the chiral structure of the electronic
% state can be characterized by a combination of two directions: rotation
% in the plane and displacement orthogonal to it (see Fig. \ref{fig:chi_p}, inset).
%Taking advantage of the simplicity of the hydrogenic chiral states,
%representing each of these classes, %and focusing on the case of aligned molecular ensembles, we gain considerable insight into 
%we illustrate
%general mechanisms that may lead to PECD in aligned chiral molecules.
%chiral effects occurring in actual chiral molecules. 

% Specifically, we aim to answer the following questions: First, which
% molecular property determines the sign of the forward-backward asymmetry
% for a given chiral molecule? Second, how does this sign depend on
% the ionization regime? Here we focus on one-photon ionization. The case of strong-field ionization will be considered in a forthcoming publication. 
Here our goal is to understand how molecular properties such as the probability density and the probability current give rise to PECD and how they affect the sign of the FBA in the one-photon ionization regime. In a forthcoming publication we will use the hydrogenic chiral wave functions to extend this study into the strong-field regime. 
% In this work we use the simplicity of the hydrogenic chiral wavefunctions to gain insight into 
% how molecular properties such as the probability density and the probability current can give rise to PECD and how they affect the sign of the FBA asymmetry. 
As a first step towards our goal, we consider the case of photoionization from a bound chiral state into an achiral Coulomb continuum, and restrict the analysis to aligned samples.

%Previous works in oriented achiral systems \cite{dubs_circular_1985-1} have demonstrated circular dichroism in the angular distribution (CDAD) within the electric-dipole approximation. However, as can be seen in Figs. 3 and 4 of the companion paper \cite{ordonez_2018_alignment}, within the electric-dipole approximation, FBA in aligned samples can only occur if the sample is chiral. The asymmetry in CDAD \cite{dubs_circular_1985-1} is with respect to the plane containing the spin of the photon and the symmetry axis of the sample, which are non-collinear. This effect is symmetry-allowed by the 

As can be seen in Fig. 2 of \cite{ordonez_generalized_2018} and in Figs. 3 and 5 of the companion paper \cite{ordonez_2018_alignment}, within the electric-dipole approximation, the photoelectron angular distribution of isotropic or aligned samples can display a FBA only if the sample is chiral. This is in contrast with other dichroic effects observed in oriented or aligned achiral systems (see e.g. \cite{dubs_circular_1985-1, ilchen_prl_2017}).

% In this sense, our work is intrinsically different from previous works in oriented (see e.g. \cite{ilchen_prl_2017}) and aligned achiral systems \cite{dubs_circular_1985-1} which display dichroism within the electric-dipole approximation.
%it was shown that photoelectron angular distributions resulting from aligned achiral bound orbitals display circular dichroism within the electric-dipole approximation \cite{dubs_circular_1985-1}, the photoelectron angular distributions display zero FBA because the total system is symmetric with respect to space inversion and the asymmetry is with respect to the plane containing the light propagation direction and the direction of molecular alignment. 

An isotropic continuum
such as that of the hydrogen atom cannot yield a FBA in an isotropically
oriented ensemble (see \cite{cherepkov1982circular} and Appendix \ref{sub:Vanishing-FBA}), because 
in this case the continuum is not able to keep track of the molecular orientations and therefore the information about the chirality of the bound state is completely washed out by the isotropic orientation averaging.
However, this does not rule out the emergence of the FBA in an aligned ensemble, where only a restricted set of orientations comes into play.
Therefore, the fact that we use an isotropic continuum shall not affect our
discussion on the origins of PECD in any way beyond what is already obvious,
namely, that the FBA we discuss relies entirely on the chirality of the bound
state and that it vanishes if we include all possible molecular orientations. 

% The approach that we developed to answer these questions is outlined in Sec. \ref{sec:a_simple_quantum_model}. 
In Sec. \ref{sub:hydrogenic_wavefunctions} we introduce the chiral hydrogenic states.
% To answer these questions we  introduce the chiral hydrogenic
% states (Sec. \ref{sub:hydrogenic_wavefunctions}). The idea of our approach is outlined in Sec. \ref{sec:a_simple_quantum_model}. 
In Sec. \ref{sec:sign_FBA_aligned}
we use the chiral hydrogenic states to
focus on physical mechanisms underlying PECD in aligned molecules. In Sec. \ref{sec:extrapolation} we discuss 
%the extension of our analysis to 
%electronic chiral states in molecules. 
effects on the FBA that result from increasing the complexity of the initial state.
In  the companion paper \cite{ordonez_2018_alignment} %Sec. \ref{sec:sign_FBA_unaligned} 
we %consider an extended picture ``complete
%physical picture'' 
show that optical propensity rules also underlie the emergence of the chiral response in photoionization in the general case of arbitrary chiral molecules and arbitrary degree of molecular alignment, and we also expose the link between the chiral response in aligned and unaligned molecular ensembles. % for results to randomly
%oriented molecules. 
Section \ref{sec:conclusions} concludes this paper. %\ref{sec:false_chiral_response}

\section{Hydrogenic chiral wave functions\label{sub:hydrogenic_wavefunctions}}

\begin{figure}%[h]
\noindent \begin{centering}
%% \includegraphics[scale=0.35]{chi_p3p1_cut_edit.png} 
%% \par\end{centering}

%% \noindent \begin{centering}
%% \includegraphics[scale=0.2]{chi_p3p1_contour_0p01_edit2.png}
\includegraphics[scale=0.35]{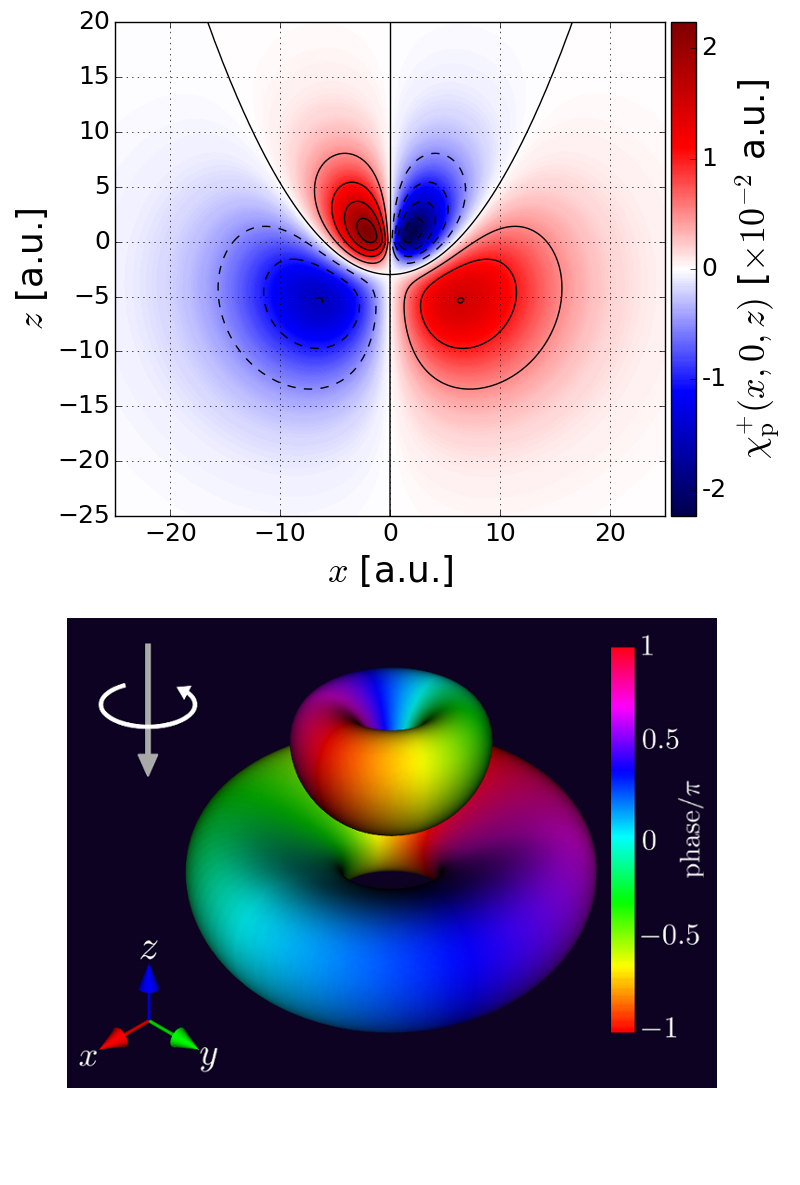} 
\par\end{centering}

\caption{Top: contour map of $\chi_{\mathrm{p}}^{+}\left(\vec{r}\right)$ {[}Eq. \eqref{eq:chi_p}{]}
on the $y=0$ plane, where it only takes real values. Dashed (solid) lines indicate negative (zero or positive) contours. Bottom: isosurface $\left|\chi_{\mathrm{p}}^{+}\left(\vec{r}\right)\right|=0.01\,\mathrm{a.u.}$
colored according to the phase. The chiral symbol on the upper left corner
indicates the polarization of the density (vertical arrow) and the
probability current in the azimuthal direction (curved arrow).\label{fig:chi_p}}
\end{figure}

We will describe three types of hydrogenic chiral wave functions.
The first type ($\mathrm{p}$-type) is of the form

\begin{equation}
\ket{\chi_{\mathrm{p}}^{\pm}}=\frac{1}{\sqrt{2}}\left(\ket{3p_{\pm1}}+\ket{3d_{\pm1}}\right),\label{eq:chi_p}
\end{equation}

where $\vert nl_{m}\rangle$ denotes a hydrogenic state with principal
quantum number $n$, angular momentum $l$, and magnetic quantum number
$m$. $\chi_{\mathrm{p}}^{+}(\vec{r})$ is shown in Fig. \ref{fig:chi_p}.
The superposition of states with even and odd values of $l$ breaks
the inversion symmetry and leads to a wave function polarized (hence
the subscript $\mathrm{p}$) along the $z$ axis, which is indicated
by an arrow pointing down in Fig. \ref{fig:chi_p}. $m=\pm1$ implies
a probability current in the azimuthal direction 
% \footnote{In general, for a wave function $\psi\left(\vec{r}\right)=\left|\psi\left(\vec{r}\right)\right|\e^{\i\xi\left(\vec{r}\right)}$
% describing a particle of mass $m$, the probability current is given
% by $\vec{j}=\frac{\hbar}{m}\left|\psi\left(\vec{r}\right)\right|^{2}\vec{\nabla}\xi\left(\vec{r}\right)$,
% which means that the gradient of the phase of the wave function in
% coordinate space indicates the direction of the probability current. }, 
and is indicated by a circular arrow in Fig. \ref{fig:chi_p}. The
combination of these two features results in a chiral wave function,
as is evident from its compound symbol. The sign of $m$ determines
the \emph{enantiomer} and, as usual, the two \emph{enantiomers} are
related to each other through a reflection; in this case, across the
$x=0$ plane, as follows from the symmetry of spherical harmonics\footnote{We could have also defined opposite enantiomers through an inversion,
and in this case instead of changing $m$ we would change the relative
sign between $\ket{3p_{1}}$ and $\ket{3d_{1}}$. Both definitions
of the opposite enantiomer are equivalent and are related to each
other via a rotation.}.

\begin{figure}
\noindent \begin{centering}
%\includegraphics[scale=0.35]{plots_paper/chi_c3p1_cut_edit}
%\includegraphics[scale=0.35]{plots_paper/chi_c3p1_cut_vel_edit} 
%\par\end{centering}

%\noindent \begin{centering}
%\includegraphics[scale=0.12]{plots_paper/chi_c3p1_contour_0p005_edit}~\includegraphics[scale=0.1%2]{plots_paper/chi_c3p1_contour_0p011_edit} 

\includegraphics[scale=0.93]{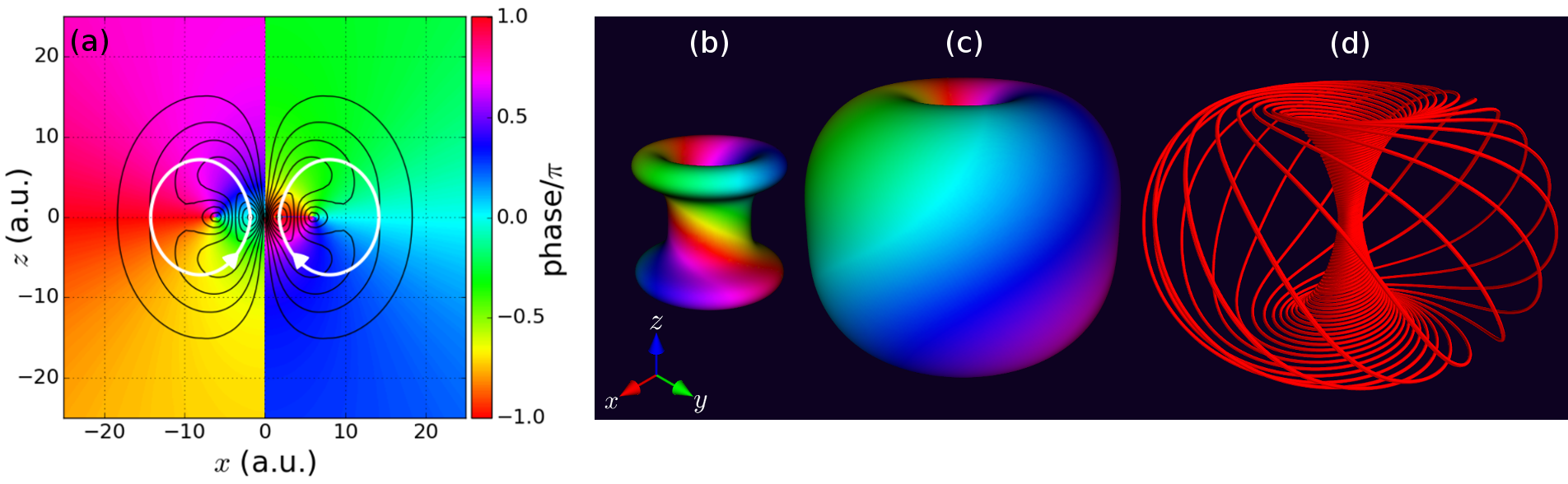}
\par\end{centering}

\caption{\textbf{(a)} Cut of $\chi_{\mathrm{c}}^{+}\left(\vec{r}\right)$ {[}Eq. \eqref{eq:chi_c}{]}
on the $y=0$ plane. The black lines indicate the contours of $\left|\chi_{\mathrm{c}}^{+}\left(\vec{r}\right)\right|$
while the colors indicate its phase. The white arrows indicate the
direction of the component of the probability current in the $y=0$ plane. \textbf{(b)}
Isosurfaces $\left|\chi_{\mathrm{c}}^{+}\left(\vec{r}\right)\right|=0.011\,\mathrm{a.u.}$
and \textbf{(c)} $\left|\chi_{\mathrm{c}}^{+}\left(\vec{r}\right)\right|=0.005\,\mathrm{a.u.}$
colored according to the phase.
\textbf{(d)} Trajectory followed by an element of the probability fluid $\left|\chi_{\mathrm{c}}^{+}\left(\vec{r}\right)\right|^{2}$. The rotation around the $z$ axis is counterclockwise. The radial distance in this specific trajectory varies between 1 and 18.5 a.u.
\label{fig:chi_c}}
\end{figure}

% \begin{figure}
% \noindent \begin{centering}
% \includegraphics[scale=0.2]{plots_paper/chi_c3p1_traj_long_edit} 
% \par\end{centering}

% \caption{Trajectory followed by an element of the probability fluid $\left|\chi_{\mathrm{c}}^{+}\left(\vec{r}\right)\right|^{2}$
% {[}Eq. \eqref{eq:chi_c}{]}. The rotation is counterclockwise around the z axis. The radial distance in this specific trajectory varies between 1 and 18.5 a.u. \label{fig:chi_c_2}}
% \end{figure}

The second type ($\mathrm{c}$-type) is given by

\begin{equation}
\ket{\chi_{\mathrm{c}}^{\pm}}=\frac{1}{\sqrt{2}}\left(\ket{3p_{\pm1}}+\i\ket{3d_{\pm1}}\right),\label{eq:chi_c}
\end{equation}

which differs from $\ket{\chi_{\mathrm{p}}}$ only in the imaginary
coefficient in front of $\ket{3d_{\pm1}}$. At first sight, since
$\braket{\vec{r}}{3p_{\pm1}}$ and $\braket{\vec{r}}{3d_{\pm1}}$
are complex functions, one would not expect important differences
between $\mathrm{p}$ and $\mathrm{c}$ states, however, as shown
in Fig. \ref{fig:chi_c}, the $\mathrm{p}$ and $\mathrm{c}$ states
are qualitatively different. We can see that instead of the polarization
along $z$, there is probability current circulating around a nodal
circle of radius $6\,\mathrm{a.u.}$ in the $z=0$ plane, as indicated
by the two circular arrows in Fig. \ref{fig:chi_c} (a). Analogously to
the $\mathrm{p}$ states, where the polarization of the probability
density is determined by the relative sign between $\ket{3p_{\pm1}}$
and $\ket{3d_{\pm1}}$, in the $\mathrm{c}$ states the direction
of the probability current is determined by the relative sign between
$\ket{3p_{\pm1}}$ and $\i\ket{3d_{\pm1}}$. This \emph{vertical }current
combined with the \emph{horizontal}\footnote{Although for simplicity we use the adjectives vertical and horizontal,
we should use instead polar and azimuthal, respectively, to be rigorous.}\emph{ }current in the azimuthal direction due to $m=\pm1$ leads
to a chiral probability current (hence the $\mathrm{c}$ subscript),
visualized in Fig. \ref{fig:chi_c} (d) via the trajectory followed
by an element of the probability fluid $\left|\chi_{\mathrm{c}}^{+}\right|^{2}$.
This single trajectory (also known as a streamline in the context
of fluids) clearly shows how, although pure helical motion of the
electron is not compatible with a bound state, helical motion can
still take place in a bound state via opposite helicities in the \emph{inner}
and \emph{outer} regions\footnote{We will say that a point is in the inner/outer region if the $z$
component of its probability current is positive/negative.}. As can be inferred from the cut of $\chi_{\mathrm{c}}^{+}\left(\vec{r}\right)$
in the $y=0$ plane [Fig. \ref{fig:chi_c} (a)], trajectories passing far from the nodal circle,
like that shown in Fig. \ref{fig:chi_c} (d), circulate faster in the
azimuthal direction than around the nodal circle while those close
to the nodal circle have the opposite behavior and look like the wire
in a toroidal solenoid. Interestingly, a probability current with the same topology was found in Ref. \cite{Hegstrom_1988} when analyzing the effect of the (chiral) weak interaction on the hydrogenic state $2p_{1/2}$.

So far we have only considered wave functions with achiral probability
densities whose chirality relies on non-zero probability currents.
The helical phase structure of $\chi_{\mathrm{c}}^{\pm}\left(\vec{r}\right)$
[see Figs. \ref{fig:chi_c} (b) and (c)] suggests that we can construct a wave
function $\chi_{\rho}^{\pm}\left(\vec{r}\right)$ with chiral probability
density (hence the subscript $\rho$) by taking the real part of $\chi_{\mathrm{c}}^{\pm}\left(\vec{r}\right)$,
i.e.

\begin{eqnarray}
\ket{\chi_{\rho}^{\pm}} & = & \frac{1}{\sqrt{2}}\left(\ket{\chi_{\mathrm{c}}^{\pm}}+\mathrm{c.c.}\right)\label{eq:chi_rho_311}\\
 & = & \frac{1}{2}\left(\ket{3p_{\pm1}}+\i\ket{3d_{\pm1}}-\ket{3p_{\mp1}}+\i\ket{3d_{\mp1}}\right)\nonumber \\
 & = & \frac{1}{\sqrt{2}}\left[\mp\ket{3p_{x}}+\ket{3d_{yz}}\right].\nonumber 
\end{eqnarray}

It turns out that this wave function is not chiral. Nevertheless,
increasing the $l$ values by one results in the wave function we
are looking for\footnote{It is also possible to obtain a chiral $\rho$ state without increasing
the value of $l$ by replacing the $\mathrm{c}$ state in Eq. \eqref{eq:chi_rho_311}
by a superposition of the $\mathrm{p}$ {[}Eq. \eqref{eq:chi_p}{]}
and $\mathrm{c}$ {[}Eq. \eqref{eq:chi_c}{]} states. However, the
resulting state is less symmetric and does not provide any more insight
than the one obtained in Eq. \eqref{eq:chi_rho} so we decided to
skip it %for now
in favor of clarity. %It will be considered in Sec. \ref{sec:extrapolation}
}. The third type ($\rho$-type) of chiral wave function is given by

\begin{eqnarray}
\ket{\chi_{\rho\left(421\right)}^{\pm}} & = & \frac{1}{\sqrt{2}}\left(\ket{\chi_{\mathrm{c}\left(421\right)}^{\pm}}+\mathrm{c.c.}\right)\label{eq:chi_rho}\\
 & = & \frac{1}{2}(\ket{4d_{\pm1}}+\i\ket{4f_{\pm1}}%
-\ket{4d_{\mp1}}+\i\ket{4f_{\mp1}})\nonumber \\
 & = & \frac{1}{\sqrt{2}}\left(\mp\ket{4d_{xz}}+\ket{4f_{yz^{2}}}\right)\nonumber 
\end{eqnarray}

\begin{figure}
\noindent \begin{centering}
%% \includegraphics[scale=0.12]{chi_rho4d1_contour_0p001_x_edit}~\includegraphics[scale=0.12]{chi_rho4d1_contour_0p008_x_edit} 
%% \par\end{centering}

%% \smallskip{}

%% \noindent \begin{centering}
%% \includegraphics[scale=0.12]{chi_rho4d1_contour_0p001_z_edit}~\includegraphics[scale=0.12]{chi_rho4d1_contour_0p008_z_edit} 
%% % \includegraphics[scale=0.15]{plots_paper/chi_rho4d1_composite2.png}
\includegraphics[scale=0.12]{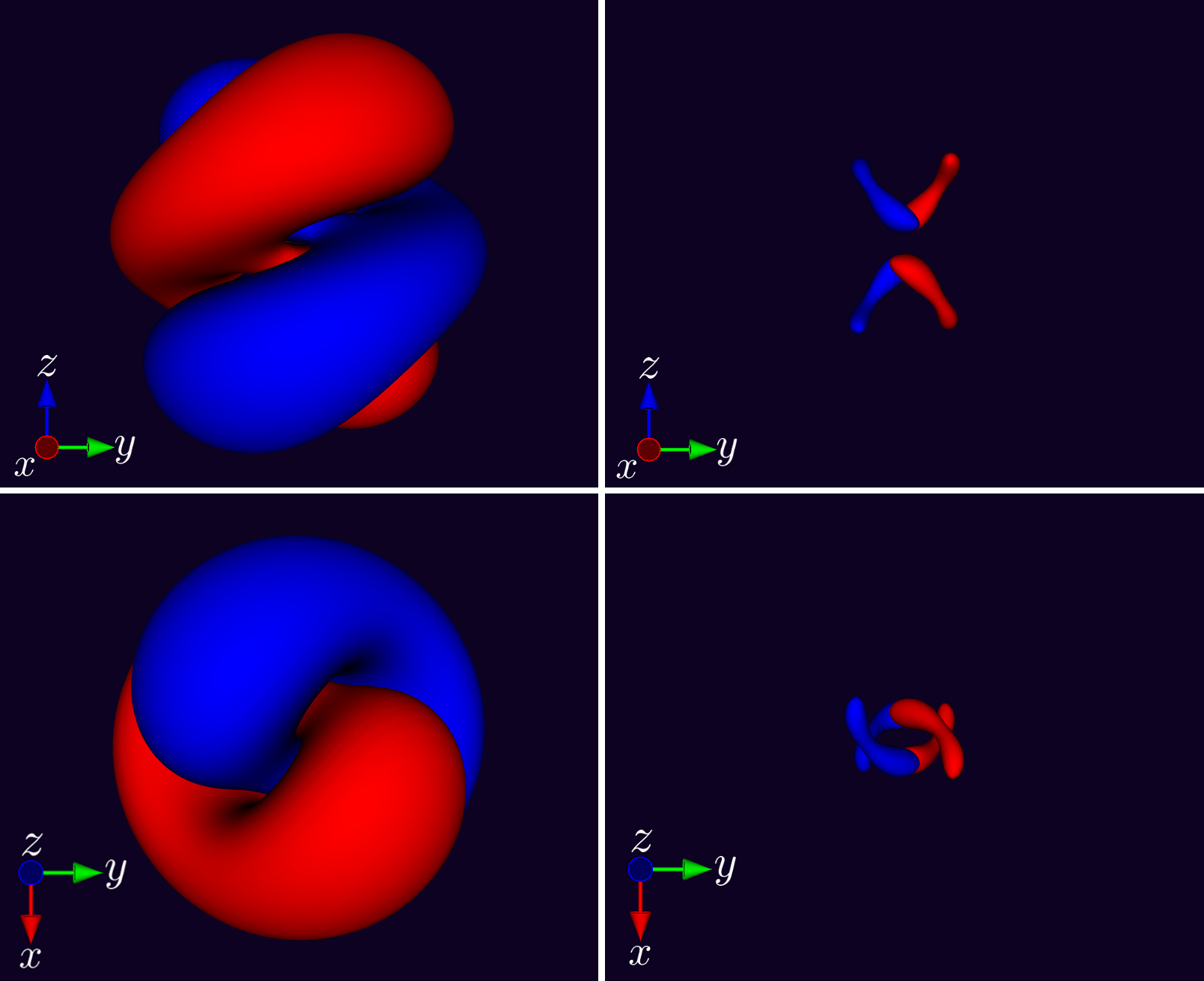}
\par\end{centering}

\caption{Isosurfaces $\chi_{\mathrm{\rho}\left(421\right)}^{+}\left(\vec{r}\right)=\pm0.001\,\mathrm{a.u.}$
(left) and $\chi_{\mathrm{\rho}\left(421\right)}^{+}\left(\vec{r}\right)=\pm0.008\,\mathrm{a.u.}$
(right) {[}Eq. \eqref{eq:chi_rho}{]} viewed along the $x$ (top)
and $z$ (bottom) axes. \label{fig:chi_rho_4d1}}
\end{figure}

\begin{figure}
\noindent \begin{centering}
%% %% \includegraphics[scale=0.35]{plots_paper/chi_c4d1_cut}
%% \includegraphics[scale=0.35]{chi_c4d1_cut_vel_edit.png} 
%% \par\end{centering}

%% \noindent \begin{centering}
%% \includegraphics[scale=0.12]{chi_c4d1_contour_0p001_edit}~\includegraphics[scale=0.12]{chi_c4d1_contour_0p004_edit}
\includegraphics[scale=0.50]{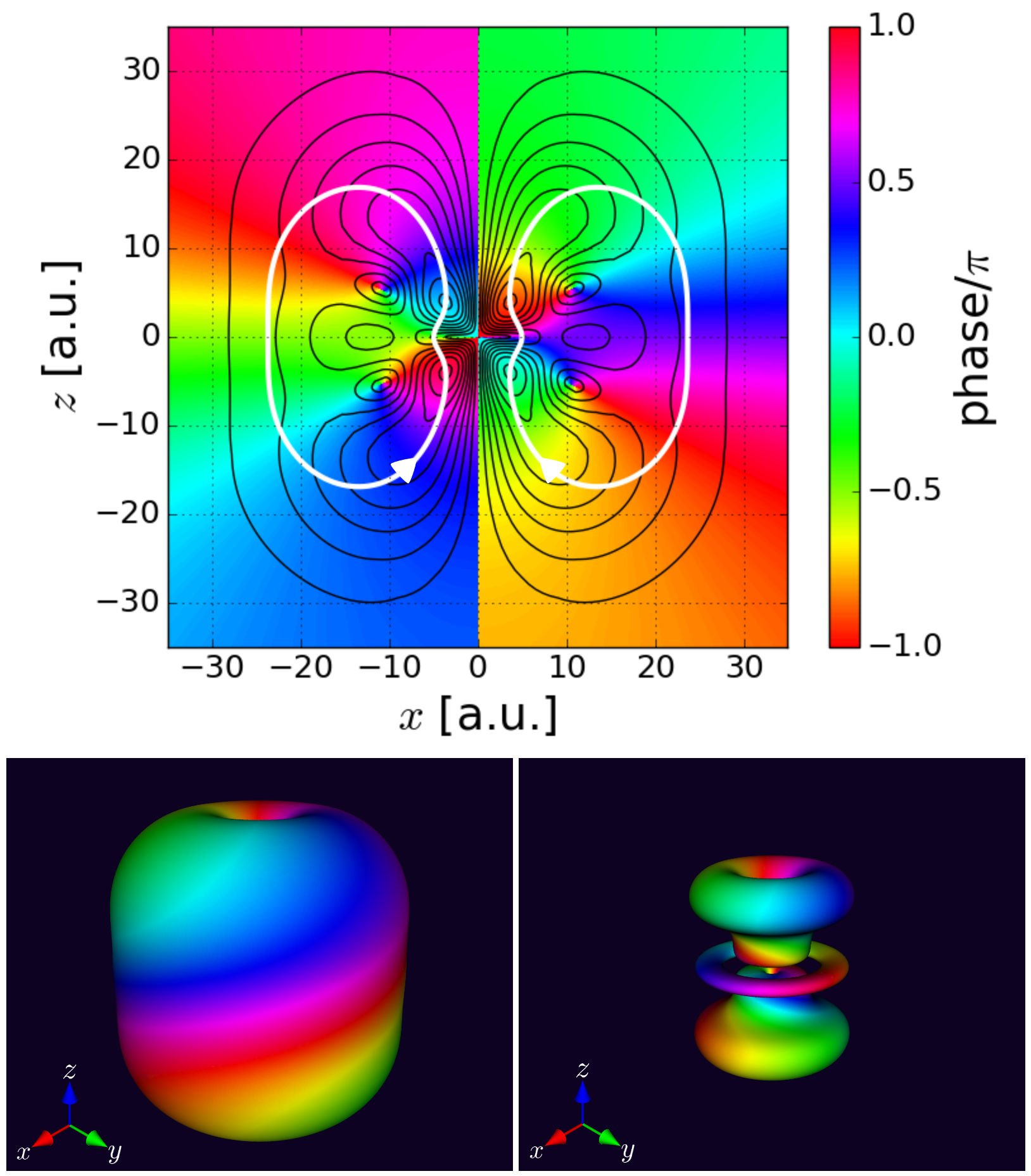}   
\par\end{centering}

\caption{Top: cut of $\chi_{\mathrm{c}\left(421\right)}^{+}\left(\vec{r}\right)$
{[}Eq. \eqref{eq:chi_c_general}{]} on the $y=0$ plane. The black
lines indicate the contours of $\vert\chi_{\mathrm{c}\left(421\right)}^{+}\left(\vec{r}\right)\vert$
while the colors indicate its phase. The white arrows indicate the
direction of the component of the probability current in the $y=0$ plane. Bottom: isosurfaces $\vert\chi_{\mathrm{c}\left(421\right)}^{+}\left(\vec{r}\right)\vert=0.001\,\mathrm{a.u.}$
(left) and $\vert\chi_{\mathrm{c}\left(421\right)}^{+}\left(\vec{r}\right)\vert=0.004\,\mathrm{a.u.}$
(right) colored according to the phase. \label{fig:chi_c4d1}}
\end{figure}
\begin{figure}
\noindent \begin{centering}
%% %\includegraphics[scale=0.35]{plots_paper/chi_c4d2_cut}
%% \includegraphics[scale=0.35]{chi_c4d2_cut_vel.png} 
%% \par\end{centering}

%% \noindent \begin{centering}
%% \includegraphics[scale=0.12]{chi_c4d2_contour_0p003_edit.png}~\includegraphics[scale=0.12]{chi_c4d2_contour_0p005_edit} 
\includegraphics[scale=0.20]{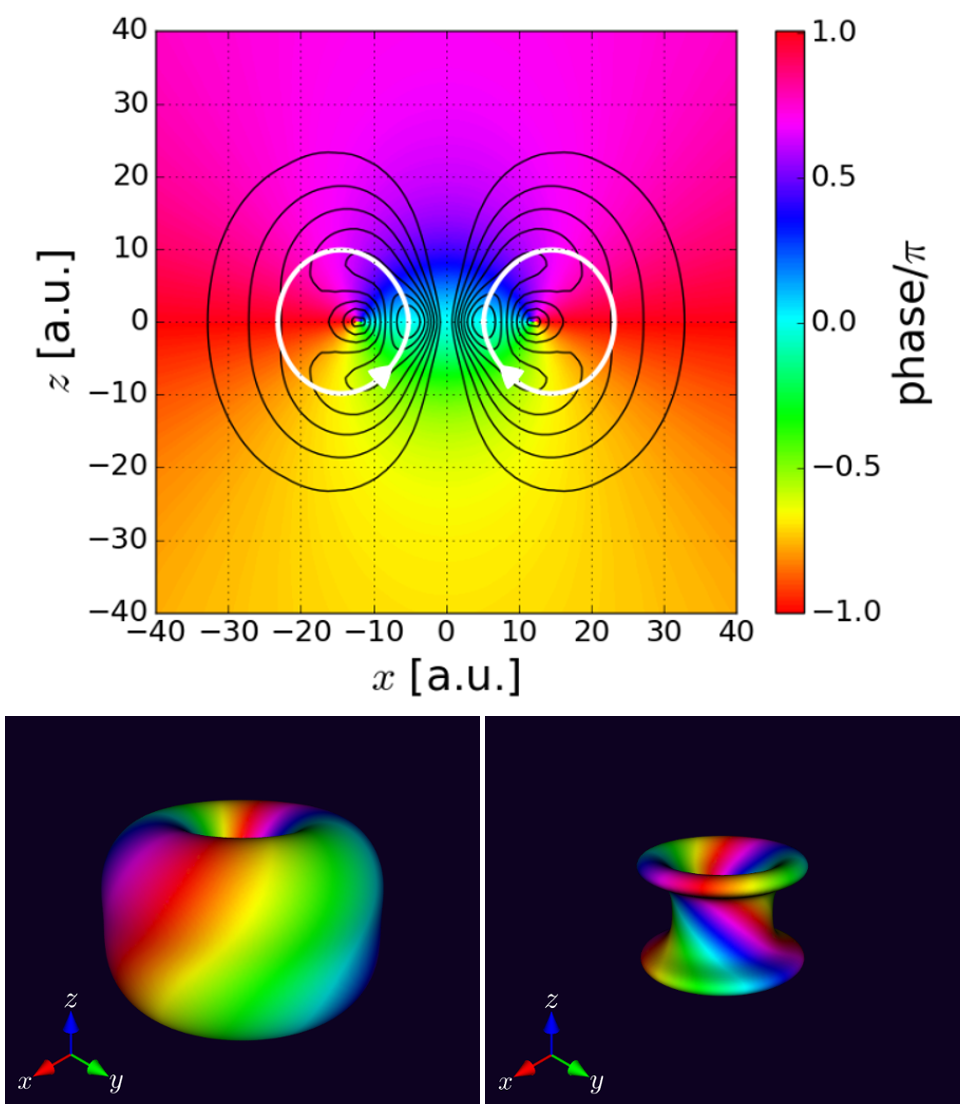}
\par\end{centering}

\caption{Top: cut of $\chi_{\mathrm{c}\left(422\right)}^{+}\left(\vec{r}\right)$
{[}Eq. \eqref{eq:chi_c_general}{]} on the $y=0$ plane. The white arrows indicate the
direction of the component of the probability current in the $y=0$ plane. Bottom: isosurfaces
$\vert\chi_{\mathrm{c}\left(422\right)}^{+}\left(\vec{r}\right)\vert=0.003\,\mathrm{a.u.}$
(left) and $\vert\chi_{\mathrm{c}\left(422\right)}^{+}\left(\vec{r}\right)\vert=0.005\,\mathrm{a.u.}$
(right) colored according to the phase. \label{fig:chi_c4d2}}
\end{figure}
\begin{figure}
\noindent \begin{centering}
%% \includegraphics[scale=0.12]{chi_rho4d2_contour_0p006_x_edit}~\includegraphics[scale=0.12]{chi_rho4d2_contour_0p001_x_edit} 
%% \par\end{centering}

%% \smallskip{}

%% \noindent \begin{centering}
%% \includegraphics[scale=0.12]{chi_rho4d2_contour_0p006_z_edit}~\includegraphics[scale=0.12]{chi_rho4d2_contour_0p001_z_edit} 
%% % \includegraphics[scale=0.15]{plots_paper/chi_rho4d2_composite.png}
\includegraphics[scale=0.12]{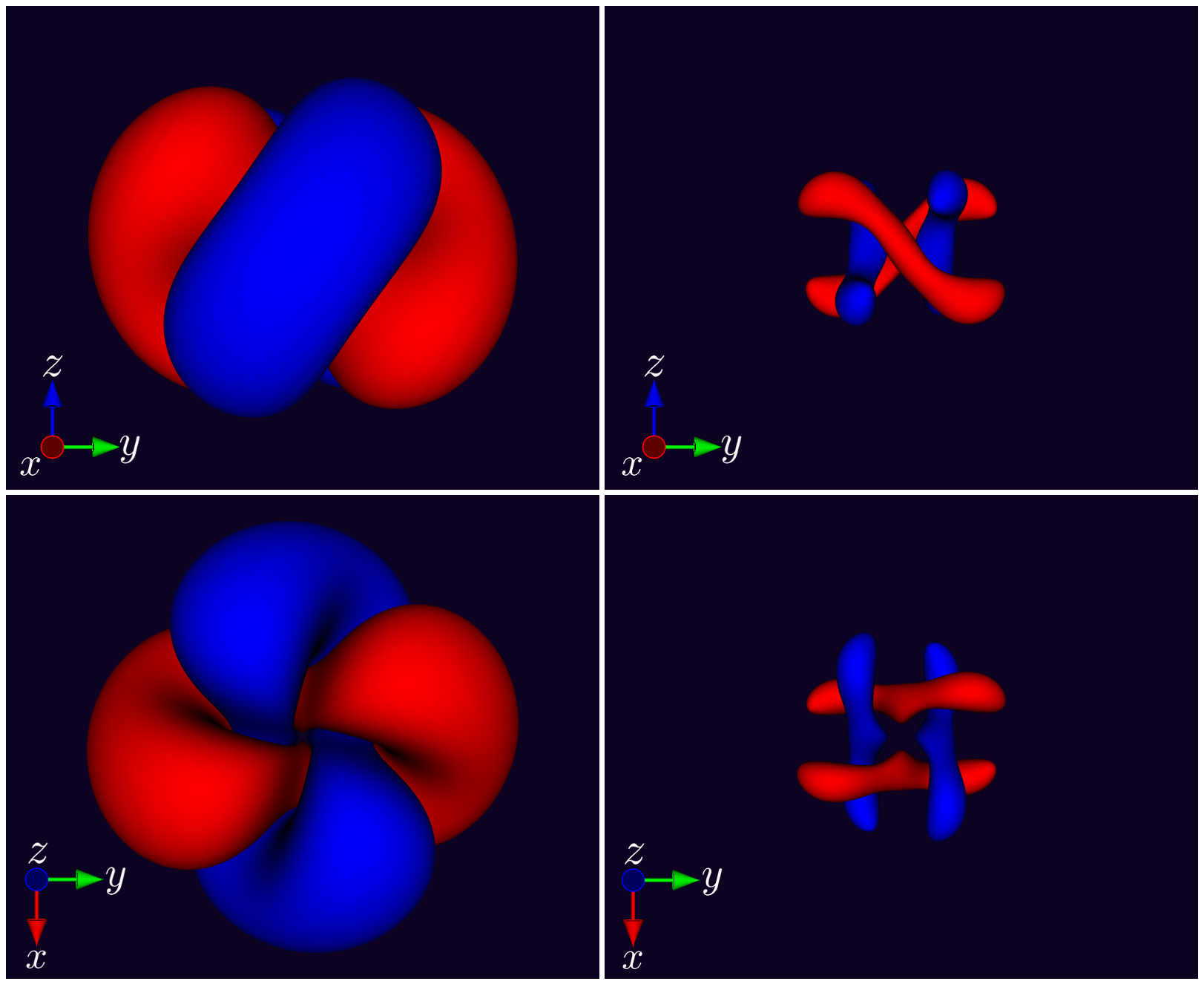}
\par\end{centering}

\caption{Isosurfaces $\chi_{\mathrm{\rho}\left(422\right)}^{+}\left(\vec{r}\right)=\pm0.001\,\mathrm{a.u.}$
(left) and $\chi_{\mathrm{\rho}\left(422\right)}^{+}\left(\vec{r}\right)=\pm0.006\,\mathrm{a.u.}$
(right) {[}Eq. \eqref{eq:chi_rho_general}{]} viewed along the $x$
(top) and $z$ (bottom) axes. \label{fig:chi_rho4d2}}
\end{figure}


and is shown in Fig. \ref{fig:chi_rho_4d1} for $m=1$. In Eq. \eqref{eq:chi_rho}
we introduced the notation

%\begin{multline}
\begin{equation}
\ket{\chi_{\mathrm{p}\left(nl\left|m\right|\right)}^{\pm}}\equiv\frac{1}{\sqrt{2}}(\ket{n,l,\pm\left|m\right|}+\ket{n,l+1,\pm\left|m\right|})\label{eq:chi_p_general}
\end{equation}
%\end{multline}

%\begin{multline}
\begin{equation}
\ket{\chi_{\mathrm{c}\left(nl\left|m\right|\right)}^{\pm}}\equiv\frac{1}{\sqrt{2}}(\ket{n,l,\pm\left|m\right|}%
+\i\ket{n,l+1,\pm\left|m\right|})\label{eq:chi_c_general}
\end{equation}
%\end{multline}

\begin{equation}
\ket{\chi_{\rho\left(nl\left|m\right|\right)}^{\pm}}\equiv\frac{1}{\sqrt{2}}\left(\ket{\chi_{\mathrm{c},\left(nl\left|m\right|\right)}^{\pm}}+\mathrm{c.c.}\right),\,l\geq2,\label{eq:chi_rho_general}
\end{equation}

which includes straightforward modifications to the simplest cases
in Eqs. \eqref{eq:chi_p}, \eqref{eq:chi_c}, and \eqref{eq:chi_rho}
that we have already considered. Figure \ref{fig:chi_c4d1} shows
$\chi_{\mathrm{c}\left(421\right)}^{+}\left(\vec{r}\right)$, which
was used in Eq. \eqref{eq:chi_rho}, and Figs. \ref{fig:chi_c4d2}
and \ref{fig:chi_rho4d2} the $m=2$ variations $\chi_{\mathrm{c}\left(422\right)}^{+}\left(\vec{r}\right)$
and $\chi_{\rho\left(422\right)}^{+}\left(\vec{r}\right)=\left(\braket{\vec{r}}{4d_{x^{2}-y^{2}}}-\braket{\vec{r}}{4f_{xyz}}\right)/\sqrt{2}$
\footnote{Interestingly, when plotted as in Fig. \ref{fig:chi_rho4d2}, the
states $\chi_{\rho\left(l+1,l,l\right)}^{\pm}\left(\vec{r}\right)$
form a topological structure known as torus link with linking number
$\pm l$.}, which will be used for the analysis of PECD in the next subsection.
As can be seen in Figs. \ref{fig:chi_rho_4d1} and \ref{fig:chi_rho4d2},
like the $\mathrm{c}$ states, the $\rho$ states also have helical
structures of opposite handedness in the \emph{inner} and \emph{outer}
regions.

The $\rho$ states are particularly meaningful because they mimic
the electronic ground state of an actual chiral molecule in the sense
that unlike the $\mathrm{p}$ and the $\mathrm{c}$ states, their
chirality is completely encoded in the probability density and does
not rely on probability currents. The decomposition of $\rho$ states
into $\mathrm{c}$ states is the chiral analogue of the decomposition
of a standing wave into two waves traveling in opposite directions,
and, as we shall see in the next subsection, it will provide the corresponding
advantages.

Finally, note that according to Barron's definition of true and false chirality \cite{barron_true_1986}, the p states display false chirality because a time reversal yields the opposite enantiomer, while the c and $\rho$ states display true chirality because a time-reversal yields the same enantiomer.

\section{The sign of the forward-backward asymmetry in aligned chiral hydrogen\label{sec:sign_FBA_aligned}}

\begin{figure}
\noindent \begin{centering}
%% \includegraphics[scale=0.35]{psi_k_0p3_abs} 
%% \par\end{centering}

%% \noindent \begin{centering}
%% ~~~~~\includegraphics[scale=0.35]{psi_k_0p3_ang}
\includegraphics[scale=0.35]{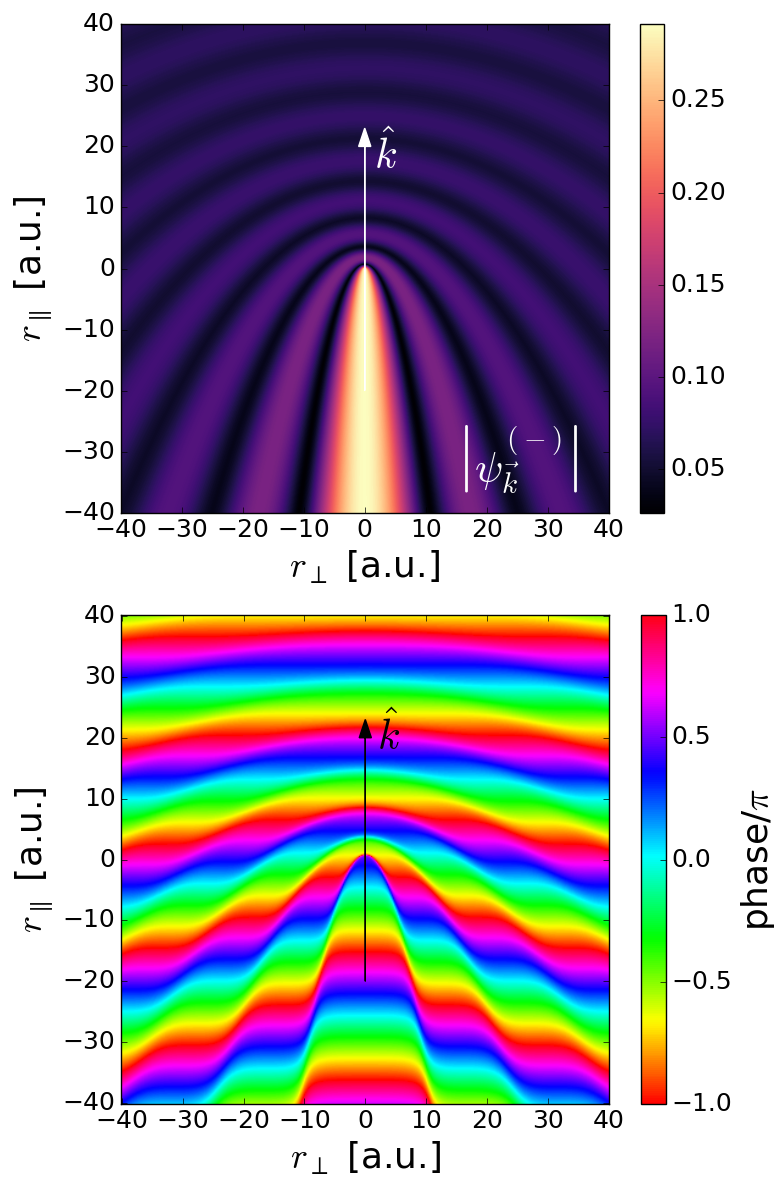} 
\par\end{centering}

\caption{Absolute value (top) and phase (bottom) of the scattering wave function
$\psi_{\vec{k}}^{\left(-\right)}$ %{[}Eq. \eqref{eq:psi_k}{]} 
evaluated
in a plane containing $\vec{k}$ for $k=0.3\,\mathrm{a.u.}$ $r_{\parallel}$
and $r_{\perp}$ are the coordinates parallel and perpendicular to
$\vec{k}$, respectively. \label{fig:psi_k}}
\end{figure}

Now we consider photoionization from the chiral bound states just
introduced via circularly polarized light. For this, we require the
scattering wave function $\psi_{\vec{k}}^{\left(-\right)}$. In the
case of hydrogen, this wave function is known analytically \cite{hans_a._bethe_quantum_1957}.
$\psi_{\vec{k}}^{\left(-\right)}\left(\vec{r}\right)$ has cylindrical
symmetry with respect to $\vec{k}$ and is shown in Fig. \ref{fig:psi_k}
for $k=0.3\,\mathrm{a.u.}$ in a plane containing $\vec{k}$. Since
only hydrogenic functions are involved, the calculation of the transition
dipole matrix element $\langle\psi_{\vec{k}}^{\left(-\right)}\vert\vec{r}\ket{\chi}$
can be carried out analytically. The angular integrals reduce to 3-j
symbols \cite{brink_angular_1968} and the radial integrals can be calculated using
the method of contour integration described in \cite{hans_a._bethe_quantum_1957}.% (see also Ref. \cite{Burgess}). 

The angle-integrated photoelectron current $\vec{j}\left(k\right)$
can be extracted from the angular and energy dependent ionization
probability $W_{\sigma}\equiv\vert\langle\psi_{\vec{k}}^{\left(-\right)}\vert\vec{r}\t\tilde{\vec{\mathcal{E}}}_{\sigma}\ket{\chi}\vert^{2}$,
where $\tilde{\vec{\mathcal{E}}}$ is the Fourier transform of the field and $\sigma=\pm 1$ indicates the rotation direction of the field (see also Ref. \cite{ordonez_generalized_2018}). First we do a partial
wave expansion of $W_{\sigma}$,

\begin{equation}
W_{\sigma}(\vec{k}) =\sum_{l,m}b_{l,m}\left(k,\sigma\right)Y_{l}^{m}(\hat{k}),
\end{equation}

and then we replace it in the expression for the $z$ component of
the angle-integrated photoelectron current,

% \begin{align}
% j_{z}\left(k,\sigma\right) & =\int\mathrm{d}\Omega_{k}W_{\sigma}(\vec{k})k_{z}\nonumber \\
%  & =\sqrt{\frac{4\pi}{3}}k\sum_{l,m}b_{l,m}\left(k,\sigma\right)\int\mathrm{d}\Omega_{k}Y_{l}^{m}(\hat{k})Y_{1}^{0}(\hat{k})\nonumber \\
%  & =\sqrt{\frac{4\pi}{3}}kb_{1,0}\left(k,\sigma\right).\label{eq:jz_PAD}
% \end{align}
\begin{equation}
j_{z}\left(k,\sigma\right)=\int\mathrm{d}\Omega_{k}W_{\sigma}(\vec{k})k_{z} =\sqrt{\frac{4\pi}{3}}kb_{1,0}\left(k,\sigma\right).\label{eq:jz_PAD}
\end{equation}

For normalization purposes, one can also consider the radial component
of the angle-integrated photoelectron current, which yields

% \begin{align}
% j_{r}\left(k,\sigma\right) & =\int\mathrm{d}\Omega_{k}W_{\sigma}(\vec{k})k\nonumber \\
%  & =\sqrt{4\pi}k\sum_{l,m}b_{l,m}\left(k,\sigma\right)\int\mathrm{d}\Omega_{k}Y_{l}^{m}(\hat{k})Y_{0}^{0}(\hat{k})\nonumber \\
%  & =\sqrt{4\pi}kb_{0,0}\left(k,\sigma\right)\label{eq:jr_PAD}
% \end{align}
\begin{equation}
j_{r}\left(k,\sigma\right)=\int\mathrm{d}\Omega_{k}W_{\sigma}(\vec{k})k=\sqrt{4\pi}kb_{0,0}\left(k,\sigma\right)\label{eq:jr_PAD}
\end{equation}

\begin{figure}
\noindent \begin{centering}
\includegraphics[scale=0.2]{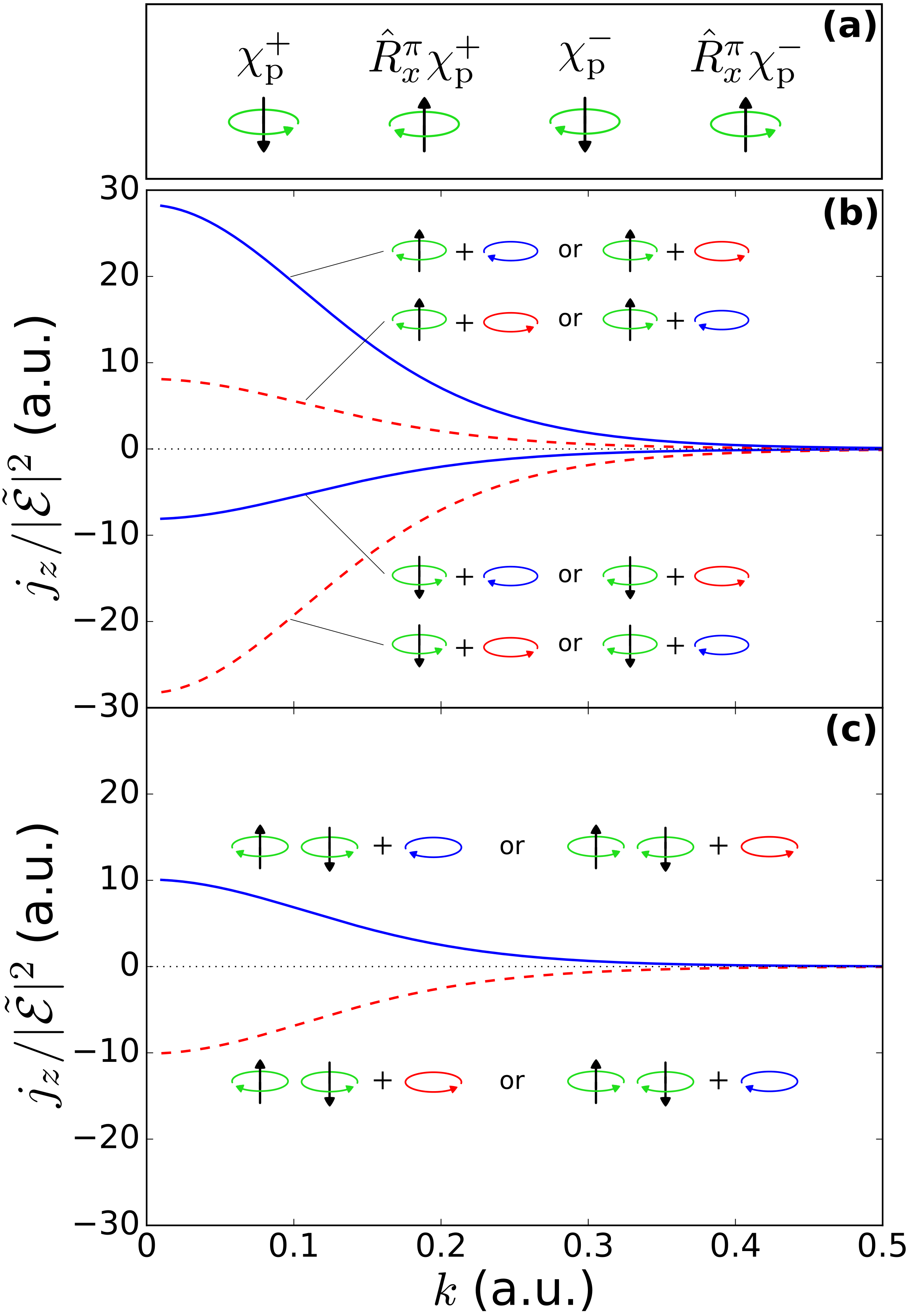}
\par\end{centering}

\caption{Photoelectron current along $z$ as a function of photoelectron momentum
resulting from photoionization of a p-type chiral state {[}see Eq.
\eqref{eq:chi_p_general}{]} via light circularly polarized in the
$xy$ plane. \textbf{(a)} Diagrams indicating the electronic polarization
(vertical arrow) and the electronic current (circular arrow) in the
p-type chiral states for two opposite enantiomers and two opposite
orientations. \textbf{$\hat{R}_{x}^{\pi}$ }is the operator that rotates
the wave function by $\pi$ radians around the $x$ axis. \textbf{(b)}
Photoelectron current {[}Eq. \eqref{eq:jz_PAD}{]} for different enantiomer,
orientation, and light polarization combinations. The enantiomer and
its orientation is indicated by the diagrams explained in (a), and
the light polarization is indicated by the circular arrows after the
plus signs. Note that the sign of $j_{z}$ is determined by the direction
of the electronic polarization and that the magnitude of\textbf{ $j_{z}$
}is determined by the relative direction between the electronic current
and the light polarization.\textbf{ (c)} Photoelectron current averaged
over two opposite orientations (equivalent to the aligned case) for
different combinations of enantiomer and light polarization. There
is no cancellation of the asymmetry because for one orientation the
bound electron co-rotates with the field, while for the opposite orientation
it counter-rotates. The calculations shown are for the states $\chi_{\mathrm{p}\left(311\right)}^{\pm}$
but the conclusions are valid for any $\chi_{\mathrm{p}\left(nlm\right)}^{\pm}$
state. \label{fig:jz_chi_p_311_summary}}

\end{figure}

\begin{figure}
\noindent \begin{centering}
\includegraphics[scale=0.3]{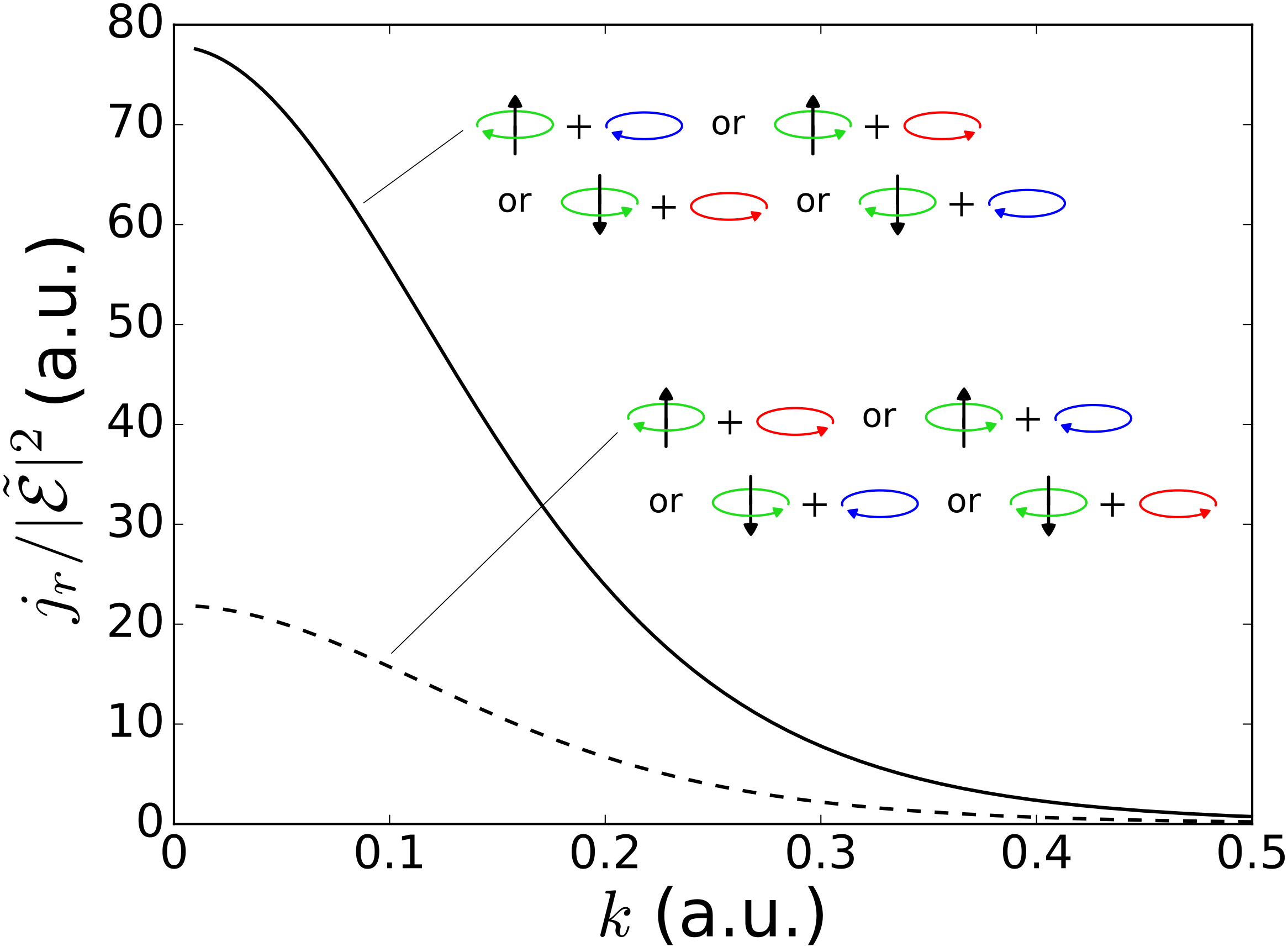}
\par\end{centering}

\caption{Total photoelectron current resulting from photoionization of a p-type
chiral state {[}see Eq. \eqref{eq:chi_p_general}{]} via light circularly
polarized in the $xy$ plane for different enantiomer, orientation,
and light polarization combinations. Diagrams are explained in Fig.
\ref{fig:jz_chi_p_311_summary} (a). Only the relative direction between
the bound electronic current and the rotating electric field determines
$j_{r}$. The calculations shown are for the states $\chi_{\mathrm{p}\left(311\right)}^{\pm}$
but the conclusions are valid for any $\chi_{\mathrm{p}\left(nlm\right)}^{\pm}$
state. \label{fig:jr_chi_p_311}}

\end{figure}

\begin{figure}
\noindent \begin{centering}
\includegraphics[scale=0.3]{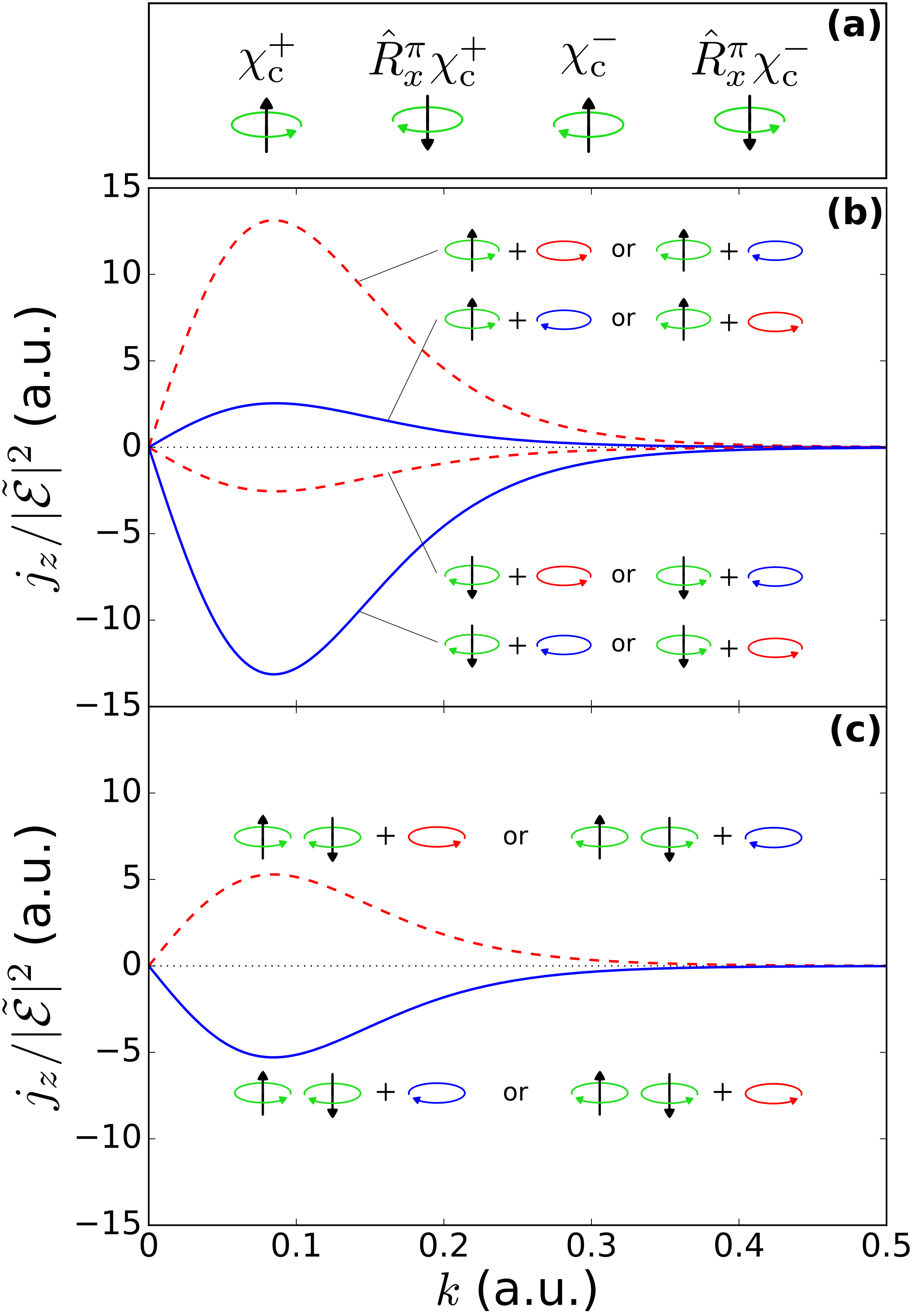}
\par\end{centering}

\caption{Same as Fig. \ref{fig:jz_chi_p_311_summary} but for the c-type
chiral states {[}see Eq. \eqref{eq:chi_c_general}{]}\textbf{. }The
role played by the electronic polarization in the p-type states is
replaced by the vertical component of the probability current in the
inner region in the c-type states. The results in (c) are also valid
for photoionization from $\rho$-type chiral states (see text). The
calculations shown are for the states $\chi_{\mathrm{c}\left(422\right)}^{\pm}$
but the conclusions are valid for any $\chi_{\mathrm{c}\left(nlm\right)}^{\pm}$
state. \label{fig:jz_chi_c_422_summary}}
\end{figure}

Figures \ref{fig:jz_chi_p_311_summary} (a), \ref{fig:jz_chi_p_311_summary} (b), and \ref{fig:jr_chi_p_311} show $j_{z}\left(k,\sigma\right)$ and $j_{r}\left(k,\sigma\right)$
for the case of photoionization from the initial states $\vert\chi_{\mathrm{p}\left(311\right)}^{\pm}\rangle$
{[}see Eq. \eqref{eq:chi_p}{]} with their $z$ molecular axis perpendicular
to the plane of polarization of the ionizing light. We can clearly see
two propensity rules that also hold for any other $\vert\chi_{\mathrm{p}\left(nlm\right)}^{\pm}\rangle$ state: 
(i) the direction
of $j_{z}$ is determined by the electronic polarization direction
of $\chi_{\mathrm{p}\left(nlm\right)}^{\pm}(\vec{r})$ and (ii) the magnitudes of
$j_{r}$ and $j_{z}$ are bigger when
the bound electron rotates in the same direction as the electric field
in comparison to when they rotate in opposite directions. The first
propensity rule is a consequence of the non-plane-wave nature of the
continuum wave function $\psi_{\vec{k}}^{\left(-\right)}\left(\vec{r}\right)$
(see Fig. \ref{fig:psi_k}), which resembles a bound polarized structure
and leads to improved overlap between $\psi_{\vec{k}}\left(\vec{r}\right)$
and $\chi_{\mathrm{p}\left(nlm\right)}^{\pm}\left(\vec{r}\right)$
in the dipole matrix element when the direction of electronic polarization
and the direction of the photoelectron coincide as compared to when
they are opposite to each other. The polarized structure of $\psi_{\vec{k}}^{\left(-\right)}\left(\vec{r}\right)$
decays monotonously with increasing $k$ and vanishes in the plane-wave
limit, which explains the monotonous decay of $j_{z}\left(k\right)$.
The second propensity rule is well known in the 1-photon-absorption atomic case \cite{hans_a._bethe_quantum_1957}. This rule changes with the ionization regime \cite{ilchen_prl_2017, barth_nonadiabatic_2011, barth_nonadiabatic_2013}.
% It
% can be either read directly from the solution of the angular integrals
% or it can be deduced from considering the overlap between spherical
% harmonics. 
% Although the numerical results presented in Figs. \ref{fig:jz_chi_p_311_summary}
% and \ref{fig:jr_chi_p_311} are specific to the states $\chi_{\mathrm{p}\left(nlm\right)}^{\pm}$
% with $\left(nlm\right)=\left(311\right)$, we have checked that they
% hold for any other $\left(nlm\right)$
% values.

In the aligned case, thanks to the vector nature of the photoelectron
current, it is enough to consider only two opposite orientations (see
Sec. III in our companion paper \cite{ordonez_2018_alignment}). In view of the first propensity
rule we have that for the two opposite orientations the polarization
will point in opposite directions and therefore $j_{z}$ will have
opposite signs. However, since for opposite orientations the bound
electron current also rotates in opposite directions while the light
polarization remains fixed, the magnitude of $j_{z}$ will be different
for each orientation, thus avoiding a complete cancellation of the
asymmetry. Furthermore, as can be seen in Fig. \ref{fig:jz_chi_p_311_summary}
(c), the sign of the orientation-averaged $j_{z}$ will be that of
the orientation where the electron co-rotates with the electric field
of the light. That is, the propensity rule for the aligned case is
that the total photoelectron current $\vec{j}=j_{z}\hat{z}$ will
point in the direction of electronic polarization associated to the
orientation where the bound electronic current co-rotates with the
ionizing electric field.

A similar analysis can be carried out for the case of photoionization
from the initial states $\vert\chi_{\mathrm{c}\left(nlm\right)}^{\pm}\rangle$,
shown in Fig. \ref{fig:jz_chi_c_422_summary} for the specific case
where $\left(nlm\right)=(422)$ but valid for any other values of $\left(nlm\right)$.
The only difference is that in this case the role which was played
by the electronic polarization in the p-type states is now played
by the vertical component of the electronic current in the inner region.
Like before, this result can be understood by considering the overlap
between the initial and final states. The polarized structure of the
continuum state determines the region contributing more to the dipole
matrix element (see $\vert\psi_{\vec{k}}^{\left(-\right)}(\vec{r})\vert$
in Fig. \ref{fig:psi_k}) and the relative direction between the probability
currents in the initial and final states in this region determines
the amount of overlap. When the direction of the probability current
of $\chi_{\mathrm{c}\left(nlm\right)}^{+}\left(\vec{r}\right)$ in
the inner region (which is where $\vert\psi_{\vec{k}}^{\left(-\right)}\left(\vec{r}\right)\vert$
is greatest) is parallel to the direction of $\vec{k}$ the overlap
is maximized. Therefore, the propensity rule in this case is that
the sign of $j_{z}$ is positive/negative when the vertical component
of the electronic current in the inner region points up/down. The
non-monotonous behavior of $j_{z}$ as a function of $k$ obeys the
fact that this propensity rule not only relies on the polarized nature
of $\vert\psi_{\vec{k}}^{\left(-\right)}\left(\vec{r}\right)\vert$,
but also on the direction of the continuum probability current, therefore,
for $k\rightarrow0$, although the density of the continuum state is maximally
polarized, its probability current tends to zero, rendering it unable
to distinguish the direction of the probability current of the bound
state, which is the feature responsible for the FBA in the first place.
At an intermediate photoelectron momentum $k\approx0.1\,\mathrm{a.u.}$
the probability current of the continuum state matches that of the
bound state and the sensitivity of the continuum state to the direction
of the probability current of the bound state is optimal. For larger
values of $k$, the match worsens and the continuum also becomes less
and less polarized leading to a monotonic decay of the FBA. The other
propensity rule regarding the relative rotation of the bound current
and the electric field remains the same and, again, the contributions
from opposite orientations to $j_{z}$ do not completely cancel each
other. 

Finally, in the case where the photoionization takes place from the states
$\vert\chi_{\rho\left(nlm\right)}^{\pm}\rangle$ {[}see Eq. \eqref{eq:chi_rho_general}{]},
there is neither any probability current nor any net polarization
that we can rely on. Furthermore, one can see from Figs. \ref{fig:chi_rho_4d1}
and \ref{fig:chi_rho4d2} that the wave function $\chi_{\rho\left(nlm\right)}^{\pm}\left(\vec{r}\right)$
is invariant with respect to rotations by $\pi$ either around the
$x$ or the $y$ axis, so that $j_{z}$ is the same for both orientations.
Thus the situation appears to be quite different from what we had
for the states $\vert\chi_{\mathrm{p}\left(nlm\right)}^{\pm}\rangle$
and $\vert\chi_{\mathrm{c}\left(nlm\right)}^{\pm}\rangle$. However,
we know that the chiral probability density of $\vert\chi_{\rho\left(nlm\right)}^{\pm}\rangle$
is the result of the superposition of the chiral currents from $\vert\chi_{\mathrm{c}\left(nlm\right)}^{\pm}\rangle$
and its complex conjugate.
% and we also know that the latter is simply
% the time-reversed version of the former. In other words, in the complex
% conjugate version the probability current flows backwards, so that
% if $\vert\chi_{\mathrm{c}\left(nlm\right)}^{\pm}\rangle$ displays
% a probability current circulating counterclockwise(+)/clockwise(-)
% around $z$, then $\vert\chi_{\mathrm{c}\left(nlm\right)}^{\pm*}\rangle$
% displays a probability current circulating clockwise(+)/counterclockwise(-)
% around $z$. And if $\vert\chi_{\mathrm{c}\left(nlm\right)}^{\pm}\rangle$
% displays a probability current moving upwards/downwards in the inner/outer
% region, then $\vert\chi_{\mathrm{c}\left(nlm\right)}^{\pm*}\rangle$
% displays a probability current moving downwards/upwards in the inner/outer
% region. 
These two chiral currents flow in opposite directions
therefore, when we subject $\vert\chi_{\rho\left(nlm\right)}^{\pm}\rangle$
to a field circularly polarized in the $xy$ plane, one part of $\vert\chi_{\rho\left(nlm\right)}^{\pm}\rangle$
will be counter-rotating and the other part will be co-rotating with
the field. One part will have an upwards vertical current in the inner
region and the other will have a downwards vertical current in the
inner region. Thus the situation for a single orientation of $\vert\chi_{\rho\left(nlm\right)}^{\pm}\rangle$
is very similar to what we had before when we considered two opposite
orientations of $\vert\chi_{\mathrm{c}\left(nlm\right)}^{\pm}\rangle$.
In fact, as shown in Appendix \ref{sub:Absence-of-m-coupling}, both
situations are exactly equivalent in the case of an isotropic continuum
like that of hydrogen. That is, the $z$ component of the photoelectron
current resulting from photoionization from the state $\vert\chi_{\rho\left(nlm\right)}^{\pm}\rangle$
is equal to that obtained from $\vert\chi_{\mathrm{c}\left(nlm\right)}^{\pm}\rangle$
after averaging over two opposite orientations. The results plotted
in Fig. \ref{fig:jz_chi_c_422_summary} (c) are not only those obtained
for $\vert\chi_{\mathrm{c}\left(422\right)}^{\pm}\rangle$, but also
those obtained for $\vert\chi_{\rho\left(422\right)}^{\pm}\rangle$.
This shows that although the $\rho$-type states do not display any
bound probability current, we can still make sense of the sign of
the FBA displayed by their photoelectron angular distribution through their decomposition into $c$-type
states.

An example of how to use these propensity rules for the less symmetric cases 
where the orientation of the molecular $z$ axis is in the plane of the 
light polarization is given in Appendix \ref{sub:propensity_in-plane}.

%\section{Extrapolation to actual chiral wave functions \label{sec:extrapolation}}
\section{Extensions of the model\label{sec:extrapolation}}

\begin{figure}
\noindent \begin{centering}
\includegraphics[scale=0.35]{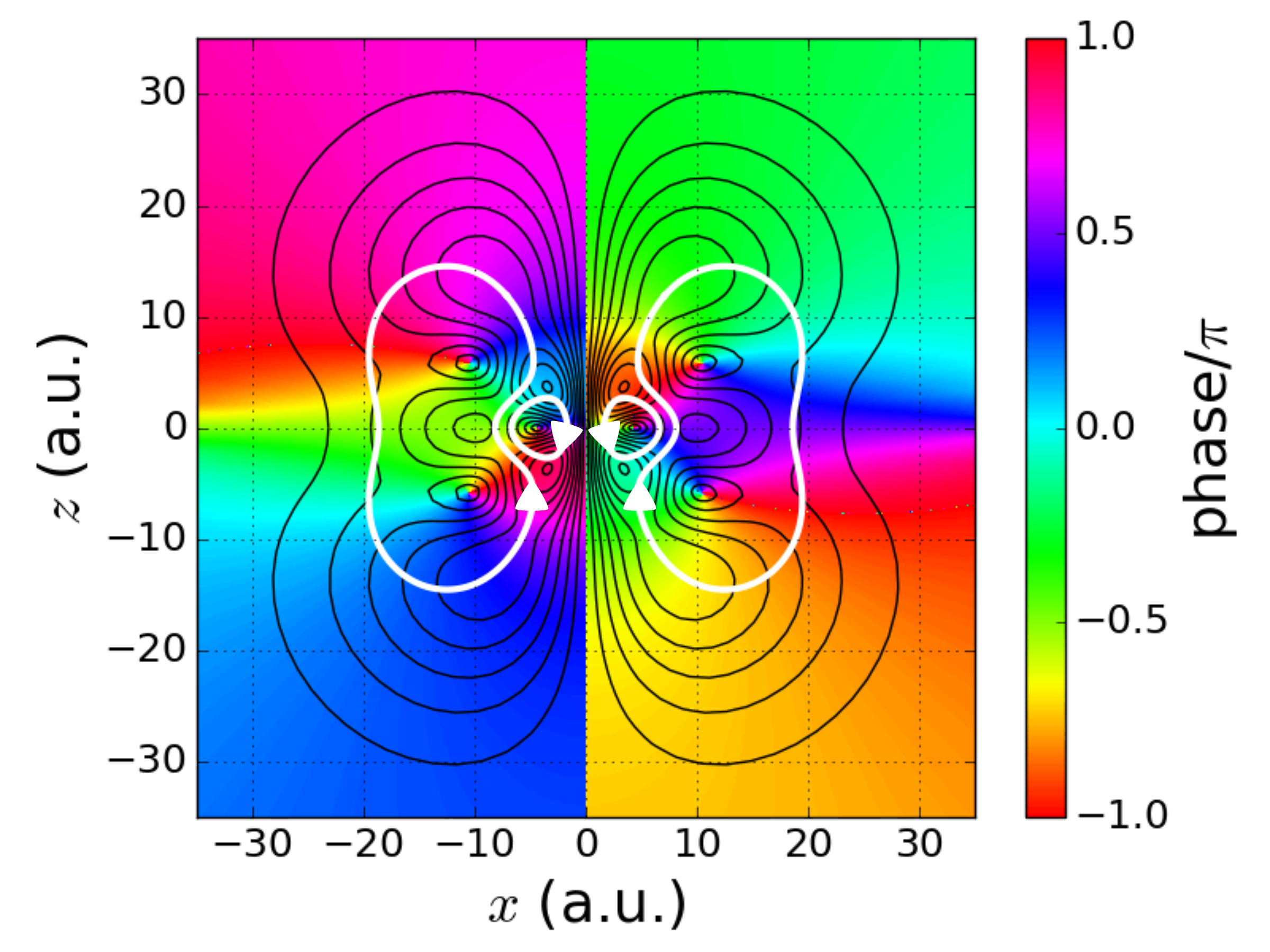}
\par\end{centering}

\caption{Cut of $\chi_{\mathrm{c}}^{+-}\left(\vec{r}\right)$
{[}Eq. \eqref{eq:chi_c4pdf1}{]} on the $y=0$ plane. The white arrows indicate the
direction of the component of the probability current in the $y=0$ plane. The extra loops in the innermost region (cf. Fig. \ref{fig:chi_c4d1}) allow for an extra degree of handedness. One handedness is associated to the big loops $(+)$ and the other to the small loops $(-)$. \label{fig:chi_c4pdf1}}
\end{figure}

\begin{figure}
\noindent \begin{centering}
\includegraphics[scale=0.5]{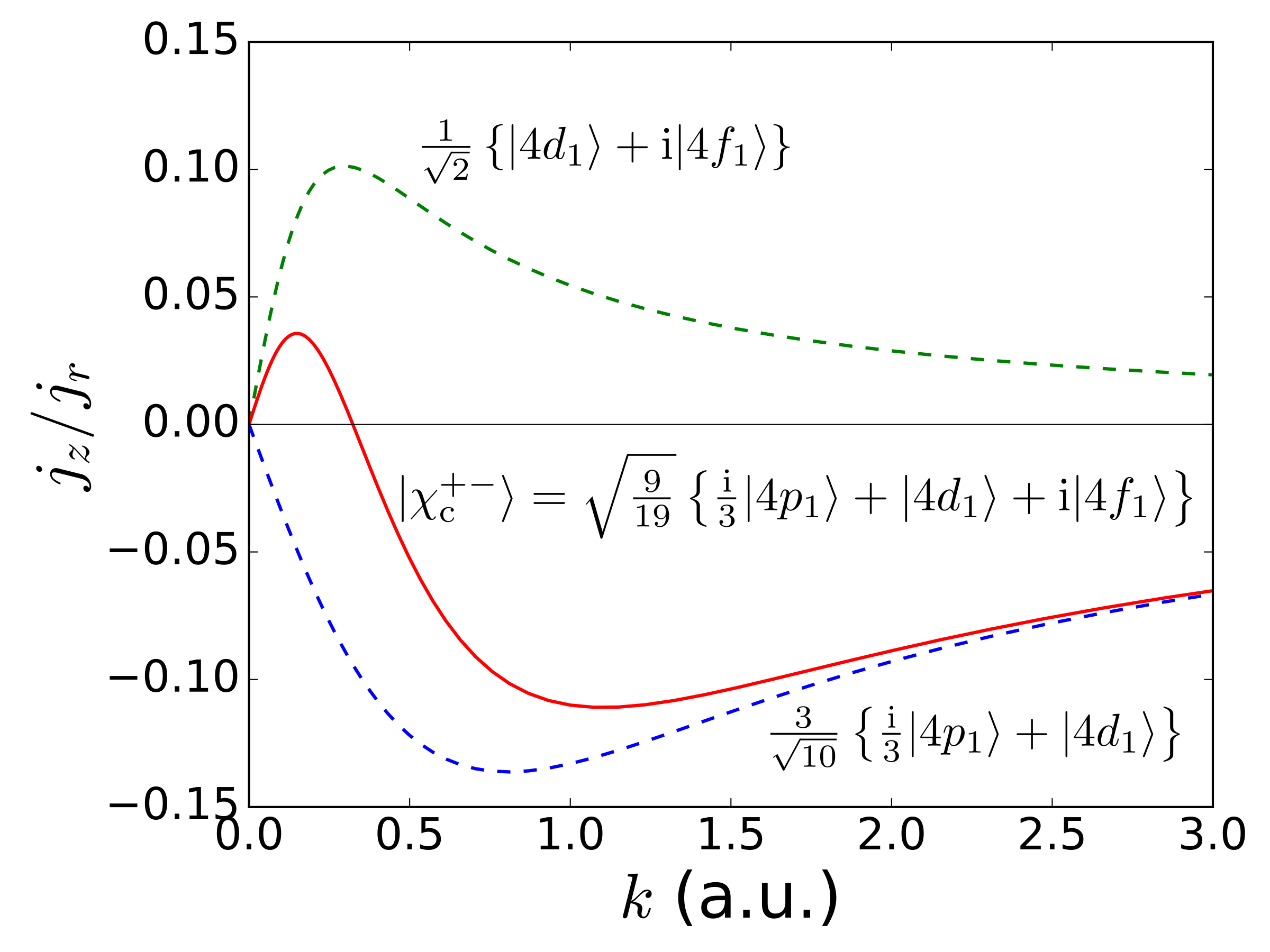}
\par\end{centering}

\caption{Normalized photoelectron current {[}Eqs. \eqref{eq:jz_PAD} and \eqref{eq:jr_PAD}{]} as a function of photoelectron momentum resulting from photoionization of the initial state $\vert \chi_{\mathrm{c}}^{+-}\rangle$ in Eq. \eqref{eq:chi_c4pdf1} (solid red line) via light left circularly polarized (rotating counter-clockwise as viewed from the $+z$ direction) and alignment perpendicular to the light polarization plane. The FBA changes sign as a function of photoelectron energy $k^2/2$ reflecting the ambivalent handedness of $\vert \chi_\mathrm{c}^{+-}\rangle$ (see Fig. \ref{fig:chi_c4pdf1}). The dashed lines show the corresponding currents for the `single-handed' states that make up $\vert \chi_{\mathrm{c}}^{+-}\rangle$.
\label{fig:jz_jr_chi_c4pdf1}}
 
\end{figure}

\begin{figure}
\noindent \begin{centering}
\includegraphics[scale=0.5]{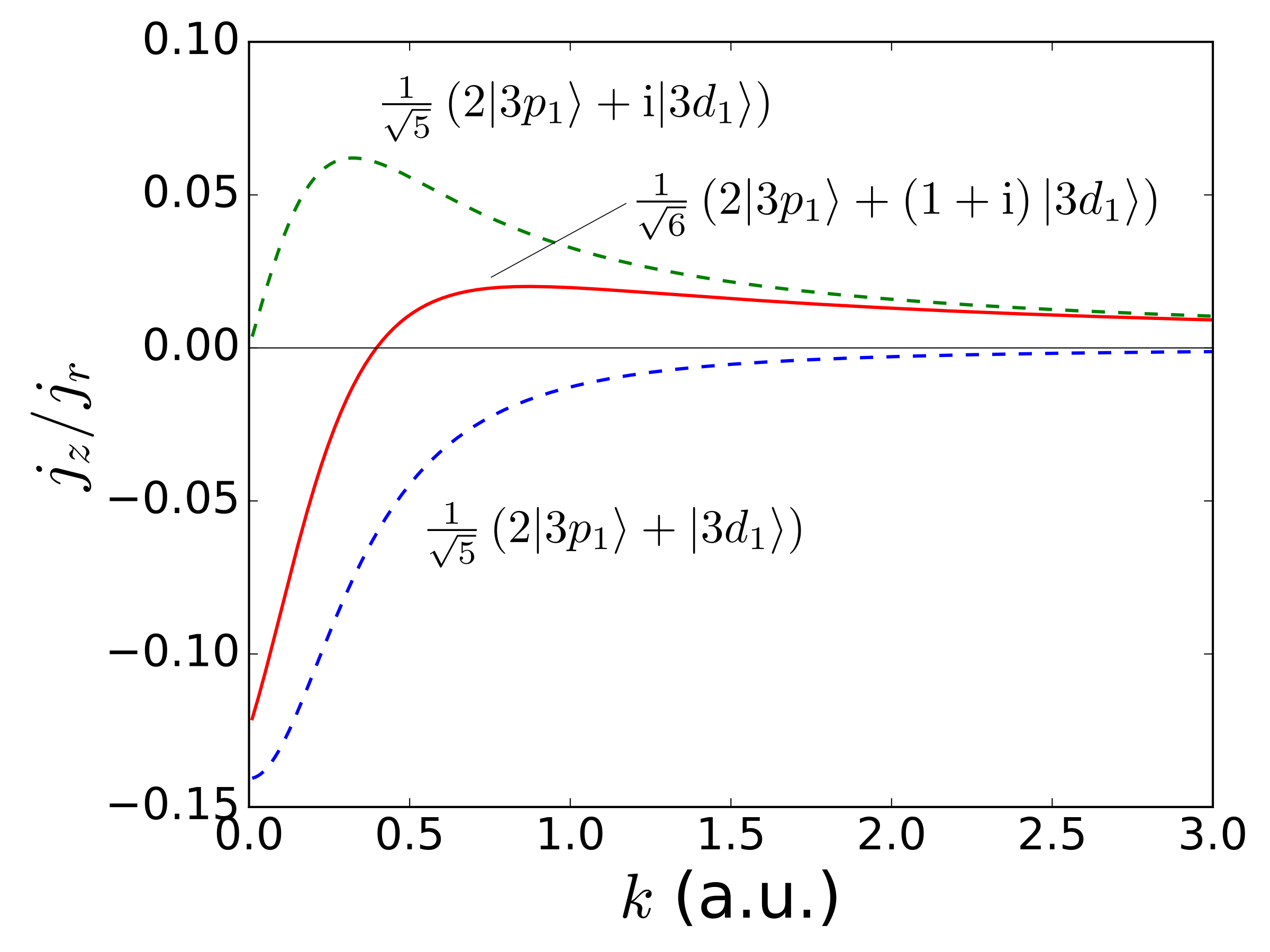}
\par\end{centering}

\caption{Normalized photoelectron current {[}Eqs. \eqref{eq:jz_PAD} and \eqref{eq:jr_PAD}{]} as a function of photoelectron momentum resulting from photoionization of the initial state $\vert \chi_{\mathrm{p}(311) }^+\rangle+\vert\chi_{\mathrm{c}(311)}^+\rangle$ (solid red line) via light left circularly polarized (rotating counter-clockwise as viewed from the $+z$ direction) and alignment perpendicular to the light polarization plane. The FBA changes sign as a function of photoelectron energy $k^2/2$ reflecting the p-like (lower dashed line) and c-like character (upper dashed line) at lower and higher photoelectron energies, respectively.\label{fig:jz_jr_chi_pc311}}
 
\end{figure}

% Our model suggests that the ground state electronic wave function
% of a real chiral molecule could, at least in principle, be treated in a way analogous to the
% $\rho$-type hydrogenic chiral state. An expansion of the wave function
% in spherical harmonics will yield a component rotating clockwise associated
% to a given FBA asymmetry, a component rotating anti-clockwise associated
% to the opposite FBA asymmetry, and a non-rotating component with its own
% FBA asymmetry. 
% The response of the component co-rotating with the field
% will be dominant over that of the counter-rotating component, and
% the asymmetry of the non-rotating component will cancel after averaging
% over two opposite orientations. However, unlike the states we considered,
% in the case of actual chiral molecules the expansion will contain
% several $l$ and $m$ values, and furthermore, the phases between
% the consecutive $l$ components will not be necessarily zero (p-type
% states) or $\pi/2$ (c-type states), but any number in between. Although
% a detailed analysis of the impact of these differences is beyond the scope
% of this work, we will briefly point out their immediate consequences.

So far we have restricted our discussion to bound wave functions involving only two different consecutive angular momenta $l$ with a specific phase between them of $0$, $\pi$ (p-type states), or $\pm\pi/2$ radians (c-type states). To get an idea of how increasing the complexity of the bound wave function may affect the FBA and the corresponding propensity rules we will consider what happens when we introduce either a third $l$ component or an arbitrary phase shift between the two $l$ components.

% Extending our model to chiral hydrogenic wave functions with more
% than two different consecutive $l$ components simply introduces the
% possibility of having a single wave function with more than one handedness
% and therefore a FBA which may change sign as a function of energy,
% a feature seen in actual molecules but absent in the simplest possible
% chiral wave functions we have presented. Consider for example the
% p-type wave function with three $l$ values $\chi_{\mathrm{p}}=w_{p}\vert4p_{1}\rangle+w_{d}\vert4d_{1}\rangle-w_{f}\vert4f_{1}\rangle$
% with $w_{p},w_{d},w_{f}>0$. On the one hand the first two terms alone
% yield polarization pointing down while the two last terms alone yield
% polarization pointing up. On the other hand the scattering wave function
% will probe differently each $l$ component of $\chi_{\mathrm{p}}$
% at different energies. Therefore, as shown in Fig. \ref{fig:jz_jr_chi_p_extra_l},
% for appropriate weight factors $w_{p}$, $w_{d}$, and $w_{f}$,
% photoionization from the state $\chi_{\mathrm{p}}$ yields a FBA which
% changes sign as a function of energy, reflecting the response from
% $w_{p}\vert4p_{1}\rangle+w_{d}\vert4d_{1}\rangle$ at higher energies
% and the response from $w_{d}\vert4d_{1}\rangle-w_{f}\vert4f_{1}\rangle$
% at lower energies. 

A third $l$ component simply introduces the
possibility of having a single wave function with more than one handedness (like a helix made of a tighter bound helix)
and therefore a FBA which may change sign as a function of energy,
a feature seen in actual molecules but absent in the simplest possible
chiral wave functions we have presented. Consider for example the
c-type wave function with three $l$ values given by

\begin{equation}
    \vert \chi_{\mathrm{c}}^{+-} \rangle = \sqrt{\frac{9}{19}}\left\{\frac{\mathrm{i}}{3}\vert4p_{1}\rangle+\vert4d_{1}\rangle+\mathrm{i}\vert4f_{1}\rangle\right\},
    \label{eq:chi_c4pdf1}
\end{equation}

which is a superposition of $\vert 4p_1 \rangle$ and the state $\vert \chi_{\mathrm{c}(421)} \rangle$ [see Eq. \eqref{eq:chi_c_general}]. A plot of this wave function on the plane $y=0$ is shown in Fig. \ref{fig:chi_c4pdf1} (compare with Fig. \ref{fig:chi_c4d1}). Unlike the states $\vert \chi_{\mathrm{c}(nml)}^{\pm}\rangle $, where the chiral current displays a single handedness, the state $\vert \chi_{\mathrm{c}}^{+-} \rangle$ displays two possible handedness, one associated with the big current loops and the other one associated with the small current loops in Fig. \ref{fig:chi_c4pdf1}. Since the two chiral currents are confined to regions of different sizes, high (low) energy photoelectrons will probe more efficiently the chirality associated to the smaller (bigger) loops, and therefore one may observe a change of sign in the FBA as the photoelectron energy is increased. Figure \ref{fig:jz_jr_chi_c4pdf1} shows how each chiral component contributes to the total FBA. An analogous behavior is observed for the case of p-type states.

Clearly, closed current loops like those shown in Fig. \ref{fig:chi_c4pdf1} can only occur around a zero of the wave function, and the emergence of the small loops in Fig. \ref{fig:chi_c4pdf1} is associated with the emergence of a zero at $r\approx4.4$ a.u., $\theta=\pi/2$. At the same time, the change of sign of the FBA is linked to the existence of the small loops, which suggests an interesting link between the topology of the wave function (zeros and currents around them) and the zeros of the FBA as a function of photoelectron energy. Further investigation of this point will be presented in a forthcoming publication. 

Introduction of phases differing from zero or $\pi/2$ between consecutive
$l$ components simply means that instead of having a pure p- or a
pure c-type state we have a superposition of both. This can also lead
to a FBA that changes sign as a function of energy because the behavior
of the FBA as a function of energy is different for p and c states.
For example, as shown in Fig. \ref{fig:jz_jr_chi_pc311}, a state
$\vert \chi_{\mathrm{p}\left(311\right)}^{+}\rangle + \vert \chi_{\mathrm{c}\left(311\right)}^{+}\rangle$
displays a FBA which is negative at lower energies and positive at
higher energies, i.e. it reflects the p character at lower energies
and the c character at higher energies.

Although the concepts of polarization, current, and wave-function overlap, underlying the propensity rules are general, the assignment of specific propensity rules to chiral molecules can be impeded due to their considerably more complex electronic structure than the elementary chiral wave-functions introduced here. In the companion paper \cite{ordonez_2018_alignment} we develop an alternative route, bypassing the specific propensity rules and introducing a more general measure, which simply indicates the presence thereof. This measure --propensity field-- controls the sign of forward-backward asymmetry in PECD.

\section{Conclusions \label{sec:conclusions}}

We have introduced three families of hydrogenic chiral wave functions
that serve as basic tools for the analysis of electronic chiral effects.
The chirality of these wave functions may be due either to a chiral
density, a chiral probability current, or a combination of achiral
density and achiral probability current.

%Since the chiral character of the wave functions stems from the appropriate combination of spherical harmonics and not from the details of the radial parts, the formalism introduced here can be extended to the analysis of chiral objects in general as long as these can be approached via partial wave expansions. 

%Furthermore, the decomposition of the chiral density state into a superposition of two counter-propagating chiral current states reveals the chiral analogue of the decomposition of a stationary plane wave into counter-propagating waves or the decomposition of a linearly polarized field into opposite circularly polarized fields. Beyond their usefulness in calculations, the value of realizing such decompositions lies in the insight they provide. 

We have used the chiral hydrogenic wave functions as a tool to explore
the basic physical mechanisms underlying the chiral response in photoionization
at the level of electrons. We have shown that two basic photoionization
propensity rules determine the sign of the forward-backward asymmetry
in photoelectron circular dichroism (PECD) in aligned molecules. %and formulated the corresponding%propensity rules that account for the forward-backward asymmetry.
One
propensity rule selects the molecular orientations in which the electron
and the electric field rotate in the same direction, and the other
propensity rule determines whether the photoelectrons are emitted
preferentially forwards or backwards. This simple picture illustrates that the  propensity rules lie at the heart of
%dynamical origin of
photoelectron circular dichroism. In the companion paper \cite{ordonez_2018_alignment} we show 
%that it holds for
how these ideas can be extended to the case of randomly oriented molecules, where another layer of effects of geometrical origin add to this simple picture.

%However, in general,
%the degree of forward-backward asymmetry arising from photoionization
%of aligned molecules can not be considered as a faithful measure of
%molecular handedness. We have shown that for a specific example of
%current carrying chiral states, but it may have broader implications,
%because aligned molecules a-priori do not allow to detect pure rotationally
%invariant signal, such as the one characterizing molecular handedness.

%The dependence of the forward-backward asymmetry as a function of photoelectron energy was studied and with the help of the propensity rules it was possible to use the plane-wave approximation to analyze the forward backward asymmetry of fast photoelectrons, which revealed that bound orbitals that differ only slightly from achiral orbitals can lead to forward-backward asymmetries even larger than those corresponding to orbitals which are clearly chiral. 

%Finally, we found that the forward-backward asymmetry depends linearly%on the degree of molecular alignment along the propagation direction%of the light, reaching its maximum magnitude for perfectly aligned%samples. 
\section{Acknowledgements}
A.F.O. and O.S. gratefully acknowledge the MEDEA project, which has received funding from the European Union's Horizon 2020 research and innovation programme under the Marie Sk\l{}odowska-Curie grant agreement 641789. O.S. acknowledges support from the DFG SPP 1840 ``Quantum Dynamics in Tailored Intense Fields'' and the DFG grant SM 292/5-2. 

\section{Appendix}

\subsection{Vanishing of the FBA for an orientation-independent continuum and
an isotropically oriented ensemble\label{sub:Vanishing-FBA}}

In this appendix we give a simple demonstration that an orientation-independent
continuum yields a zero FBA when all molecular orientations are equally
likely (see also \cite{cherepkov1982circular}). Consider the lab-frame orientation-averaged
photoelectron angular distribution

\begin{equation}
W_{\sigma}(\vec{k})=\int\mathrm{d}\lambda\left|\braoket{\psi_{\vec{k}}^{\left(-\right)}}{\hat{\epsilon}_{\sigma}}{\hat{D}(\lambda)\chi}\right|^{2},
\end{equation}

where $\hat{\epsilon}_{\sigma}\equiv\left(\hat{x}\pm\i\hat{y}\right)$,
and $\hat{D}\left(\lambda\right)$ is the operator that rotates the
bound wave function $\chi(\vec{r})$ by the Euler angles $\lambda\equiv\alpha\beta\gamma$.
We assumed that the scattering wave function $\psi_{\vec{k}}^{\left(-\right)}(\vec{r})$
is independent of the molecular orientation $\lambda$ and therefore
there is no need to rotate it. Here we consider rotations in the active
sense, i.e. we always have the same frame of reference (the lab frame)
and we rotate the functions. If we expand the bound wave function
in spherical harmonics as

\begin{equation}
\chi\left(\vec{r}\right)=\sum_{l,m}u_{l,m}\left(r\right)Y_{l}^{m}\left(\hat{r}\right),
\end{equation}

then the rotation operator $\hat{D}\left(\lambda\right)$ acts on
$\chi\left(\vec{r}\right)$ through the Wigner D-matrices $D_{m^{\prime},m}^{\left(l\right)}\left(\lambda\right)$
according to 

\begin{equation}
\hat{D}\left(\lambda\right)\chi\left(\vec{r}\right)=\sum_{l,m,m^{\prime}}D_{m^{\prime},m}^{\left(l\right)}\left(\lambda\right)u_{l,m}\left(r\right)Y_{l}^{m^{\prime}}\left(\hat{r}\right)=\sum_{l,m,m^{\prime}}D_{m^{\prime},m}^{\left(l\right)}\left(\lambda\right)\chi_{l,m,m^{\prime}}\left(\vec{r}\right).
\end{equation}

Replacing this expansion in the expression for the photoelectron angular distribution we obtain 

\begin{align}
W_{\sigma}(\vec{k}) 
% & =\int\mathrm{d}\lambda\left|\bra{\psi_{\vec{k}}^{\left(-\right)}}\hat{\epsilon}_{\sigma}\ket{\hat{D}\left(\lambda\right)\chi}\right|^{2}\nonumber \\
 & =\sum_{l_{1},m_{1},m_{1}^{\prime},l_{2},m_{2},m_{2}^{\prime}}\left[\int\mathrm{d}\lambda D_{m_{2}^{\prime},m_{2}}^{\left(l_{2}\right)*}\left(\lambda\right)D_{m_{1}^{\prime},m_{1}}^{\left(l_{1}\right)}\left(\lambda\right)\right]\nonumber \\
 & \times\braoket{\psi_{\vec{k}}^{\left(-\right)}}{\hat{\epsilon}_{\sigma}}{\chi_{l_{2},m_{2},m_{2}^{\prime}}}^{*}\braoket{\psi_{\vec{k}}^{\left(-\right)}}{\hat{\epsilon}_{\sigma}}{\chi_{l_{1},m_{1},m_{1}^{\prime}}}.\nonumber \\
 & =\sum_{l_{1},m_{1},m_{1}^{\prime}}\frac{8\pi^{2}}{2l_{1}+1}\left|\braoket{\psi_{\vec{k}}^{\left(-\right)}}{\hat{\epsilon}_{\sigma}}{\chi_{l_{1},m_{1},m_{1}^{\prime}}}\right|^{2}\label{eq:W_expansion_1}
\end{align}

where we used the orthogonality relation for the Wigner D-matrices \cite{brink_angular_1968}. Now we expand the scattering wave function in spherical
harmonics with respect to $\hat{k}$

\begin{equation}
\psi_{\vec{k}}^{\left(-\right)}\left(\vec{r}\right)=\sum_{l,m}\psi_{k,l,m}\left(\vec{r}\right)Y_{l}^{m*}(\hat{k}),
\end{equation}

and replace it in the expression for the photoelectron angular distribution

\begin{align}
W_{\sigma}(\vec{k}) & =\sum_{l,m,l_{1},m_{1},m_{1}^{\prime}}\frac{8\pi^{2}}{2l_{1}+1}\left|\braoket{\psi_{k,l,m}}{\hat{\epsilon}_{\sigma}}{\chi_{l_{1},m_{1},m_{1}^{\prime}}}Y_{l}^{m}(\hat{k})\right|^{2}\nonumber \\
 & =\sum_{l,m}f_{\sigma,l,m}\left(k\right)\left|Y_{l}^{m}(\hat{k})\right|^{2}\label{eq:W_expansion_2}
\end{align}

where 

\begin{equation}
f_{\sigma,l,m}(k)=\sum_{l_{1},m_{1},m_{1}^{\prime}}\frac{8\pi^{2}}{2l_{1}+1}\left|\braoket{\psi_{k,l,m}}{\hat{\epsilon}_{\sigma}}{\chi_{l_{1},m_{1},m_{1}^{\prime}}}\right|^{2}.
\end{equation}

Since $\vert Y_{l}^{m}(\hat{k})\vert^{2}$ is symmetric with respect
to the $xy$ plane for every $l$ and $m$, Eq. \eqref{eq:W_expansion_2}
shows that $W_{\sigma}(\vec{k})$ is also symmetric with respect to
the $xy$ plane, and thus exhibits no FBA, irregardless of the values
of the coefficients $f_{\sigma,l,m}(k)$ which encode the information
about the chiral bound state and the light polarization. Any deviation
from an orientation-independent scattering wave function will introduce
cross-terms in Eqs. \eqref{eq:W_expansion_1} and \eqref{eq:W_expansion_2},
and therefore will open the possibility of non-zero FBA.

\subsection{Absence of m-coupling in the photoelectron current for isotropic
continua\label{sub:Absence-of-m-coupling}}

Consider the photoelectron angular distribution resulting from a single molecular orientation

\begin{equation}
W_{\sigma}(\vec{k})=\left|\braoket{\psi_{\vec{k}}^{\left(-\right)}}{\hat{\epsilon}_{\sigma}}{\chi}\right|^{2},
\end{equation}

where $\hat{\epsilon}_{\sigma}\equiv\left(\hat{x}\pm\i\hat{y}\right)$,
$\chi$ is the bound wave function that has already been rotated by
the Euler angles $\lambda\equiv\alpha\beta\gamma$, and the scattering
wave function is molecular-orientation independent, i.e. it only depends
on the relative direction between the position vector $\vec{r}$ and
the photoelectron momentum $\vec{k}$. Both wave functions can be
expanded as

\begin{equation}
\chi\left(\vec{r}\right)=\sum_{l,m}\chi_{l,m}\left(\vec{r}\right)
\end{equation}

\begin{equation}
\psi_{\vec{k}}^{\left(-\right)}\left(\vec{r}\right)=\sum_{l,m}\psi_{l,m}\left(k,\vec{r}\right)Y_{l}^{m*}(\hat{k}),
\end{equation}

where $\chi_{l,m}\left(\vec{r}\right)=u_{l,m}\left(r\right)Y_{l}^{m}\left(\hat{r}\right)$
and $\psi_{l,m}\left(k,\vec{r}\right)=v_{l,m}\left(k,r\right)Y_{l}^{m}\left(\hat{r}\right)$.
Replacing these expansions in $W_{\sigma}(\vec{k})$ we get

\begin{multline}
W_{\sigma}(\vec{k})=\sum_{l_{1},m_{1},l_{2},m_{2},l_{1}^{\prime},l_{2}^{\prime}}\braoket{\psi_{l_{1}^{\prime},m_{1}+\sigma}}{\hat{\epsilon}_{\sigma}}{\chi_{l_{1},m_{1}}}^{*}\braoket{\psi_{l_{2}^{\prime},m_{2}+\sigma}}{\hat{\epsilon}_{\sigma}}{\chi_{l_{2},m_{2}}}\label{eq:W_expansion_3}\\
\times Y_{l_{1}^{\prime}}^{m_{1}+\sigma*}(\hat{k})Y_{l_{2}^{\prime}}^{m_{2}+\sigma}(\hat{k})
\end{multline}

where we used the selection rules $m_{1}^{\prime}=m_{1}+\sigma$ and
$m_{2}^{\prime}=m_{2}+\sigma$ for the electric-dipole transitions. The product
of the spherical harmonics $Y_{l_{1}^{\prime}}^{m_{1}+\sigma*}Y_{l_{2}^{\prime}}^{m_{2}+\sigma}$
can be rewritten as a superposition of spherical harmonics $Y_{l}^{m}$
with $m=-m_{1}+m_{2}$, and the calculation of $j_{z}$ only requires
the term $l,m=1,0$ [see Eq. \eqref{eq:jz_PAD}]. Therefore we must
only consider the terms in Eq. \eqref{eq:W_expansion_3} where $m_{1}=m_{2}$,
which means that the different $m$ components in the bound wave function
$\chi$ do not interfere in $j_{z}$. That is, the calculation of
$j_{z}$ for a coherent superposition $\chi_{l_{1},m_{1}}+\chi_{l_{2},m_{2}}$
yields the same result as the sum of the $j_{z}$'s obtained for each
state of the superposition separately.

\subsection{An example of propensity rules for the in-plane orientation.\label{sub:propensity_in-plane}}

Consider the state $\vert\chi_{\mathrm{p}\left(311\right)}^{+}\rangle$
when the molecular frame is related to the lab frame by a rotation
of $\pi/2$ around $\hat{y}^{\mathrm{L}}$. In this case, the electronic polarization
points along $-\hat{x}^{\mathrm{L}}$ and the bound probability current
is in the $\hat{y}^{\mathrm{L}}\hat{z}^{\mathrm{L}}$ plane. For light
circularly polarized in the $\hat{x}^{\mathrm{L}}\hat{y}^{\mathrm{L}}$
plane, neither the asymmetry of the initial state (i.e. its electronic
polarization) is along the direction perpendicular to the light polarization
nor the bound probability current is in the plane of the light polarization.
Nevertheless, with the help of the Wigner rotation matrices \cite{brink_angular_1968}
we can write the rotated spherical harmonics in terms of unrotated
spherical harmonics as

\begin{equation}
\hat{R}_{y}^{\pi/2}Y_{1}^{1}=\frac{1}{2}Y_{1}^{1}+\frac{1}{\sqrt{2}}Y_{1}^{0}+\frac{1}{2}Y_{1}^{-1},\label{eq:RyY11}
\end{equation}

\begin{equation}
\hat{R}_{y}^{\pi/2}Y_{2}^{1}=-\frac{1}{2}Y_{2}^{2}-\frac{1}{2}Y_{2}^{1}+\frac{1}{2}Y_{2}^{-1}+\frac{1}{2}Y_{2}^{-2}.\label{eq:RyY21}
\end{equation}

Replacing Eqs. \eqref{eq:RyY11} and \eqref{eq:RyY21} in the expression
for $\chi_{\mathrm{p}\left(311\right)}^{+}$ {[}Eq. \eqref{eq:chi_p}{]}
and using $\hat{R}_{y}^{\pi}Y_{l}^{m}=\left(-1\right)^{l+m}Y_{l}^{-m}$
we obtain

\begin{align}
\hat{R}_{y}^{\pi/2}\chi_{\mathrm{p}\left(311\right)}^{+} & =\frac{1}{2}\frac{1}{\sqrt{2}}\left[R_{3,1}Y_{1}^{1}-R_{3,2}Y_{2}^{1}-R_{3,2}Y_{2}^{2}\right]\nonumber \\
 & +\frac{1}{2}\frac{1}{\sqrt{2}}\left[R_{3,1}Y_{1}^{-1}+R_{3,2}Y_{2}^{-1}+R_{3,2}Y_{2}^{-2}\right]\nonumber \\
 & +\frac{1}{2}R_{3,1}Y_{1}^{0}\nonumber \\
 & =\frac{1}{2}\hat{R}_{y}^{\pi}\chi_{\mathrm{p}\left(311\right)}^{-}+\frac{1}{2}\chi_{\mathrm{p}\left(311\right)}^{-}+\frac{1}{\sqrt{2}}\Phi\label{eq:chi_p_rot}
\end{align}

where $R_{n,l}\left(r\right)$ are the bound radial functions of hydrogen
and we defined 

\begin{equation}
\Phi\left(\vec{r}\right)=\frac{1}{\sqrt{2}}\left\{ R_{3,1}\left(r\right)Y_{1}^{0}\left(\hat{r}\right)+\frac{1}{\sqrt{2}}R_{3,2}\left(r\right)\left[Y_{2}^{-2}\left(\hat{r}\right)-Y_{2}^{2}\left(\hat{r}\right)\right]\right\} .
\end{equation}

In analogy to what we did before with the $\rho$ state, we separated
the wave function according to the direction of its probability current
with respect to the $z$ axis, i.e. into positive, negative, and zero
$m$'s. In general, this is as far as we can go with the simplification,
and at this point we must only figure out the sign of the asymmetry
that the part co-rotating with the electric field yields to tell the
sign of the FBA asymmetry that the full wave function yields. However,
in this particularly simple case we can recognize that not only the
$m=0$ but also the $m=\pm2$ terms do not contribute to the chirality
of neither the co-rotating nor the counter-rotating parts. We have
grouped this achiral terms into $\Phi$. Furthermore, the remaining
terms can be rewritten in terms of p states with their polarizations
pointing along $\hat{z}^{\mathrm{L}}$ and $-\hat{z}^{\mathrm{L}}$.

From the discussion of the propensity rules in Sec. \ref{sec:sign_FBA_aligned}
and from Fig. \ref{fig:jz_chi_p_311_summary} we already know the
$j_{z}^{\mathrm{L}}$ that will result from each of the p states appearing
in Eq. \eqref{eq:chi_p_rot}. Furthermore, from Appendix \ref{sub:Absence-of-m-coupling}
we know that each $m$ component will have an independent effect on
$j_{z}$. Therefore, although the unrotated state $\chi_{\mathrm{p}\left(311\right)}^{+}\left(\vec{r}^{\mathrm{L}}\right)$
exhibits a negative $j_{z}^{\mathrm{L}}$ for both left and right
circularly polarized light, once we rotate this state by $\pi/2$
around $\hat{y}^{\mathrm{L}}$ Eq. \eqref{eq:chi_p_rot} shows that
it will exhibit a negative/positive $j_{z}^{\mathrm{L}}$ for right/left
circularly polarized light because the signal from the second/first
term will dominate.

We can also use Eq. \eqref{eq:chi_p_rot} to verify that $j_{z}$
vanishes in the isotropically-oriented case {[}see Eq. (17) %\eqref{eq:j_1_field_6_orientations}
of the companion paper \cite{ordonez_2018_alignment}{]}
by taking into account the 6 orientations of $\vert\chi_{\mathrm{p}\left(311\right)}^{+}\rangle$
displayed in Fig. 2 of \cite{ordonez_2018_alignment} %\eqref{fig:j_6_orientations_1_field}

\begin{align}
j_{z} & =\frac{1}{6}\left[j_{z}\left(\chi_{\mathrm{p}}^{+}\right)+j_{z}\left(\hat{R}_{y}^{\pi}\chi_{\mathrm{p}}^{+}\right)+4j_{z}\left(\hat{R}_{y}^{\pi/2}\chi_{\mathrm{p}}^{+}\right)\right]\label{eq:jz_iso}\\
 & =\frac{1}{6}\left[j_{z}\left(\chi_{\mathrm{p}}^{+}\right)+j_{z}\left(\hat{R}_{y}^{\pi}\chi_{\mathrm{p}}^{+}\right)+j_{z}\left(\hat{R}_{y}^{\pi}\chi_{\mathrm{p}}^{-}\right)+j_{z}\left(\chi_{\mathrm{p}}^{-}\right)\right]\nonumber \\
 & =0\nonumber 
\end{align}

where the arguments of $j_{z}$ on the right hand side of Eq. \eqref{eq:jz_iso}
indicate the orientation of the initial wave function, and from symmetry
we know that the four orientations where $\hat{z}^{\mathrm{M}}$ lies
on the $\hat{x}^{\mathrm{L}}\hat{y}^{\mathrm{L}}$ plane yield the
same photoelectron current.

\bibliographystyle{apsrev4-1}
\bibliography{MyLibrary}

\end{document}